\documentclass[prd,preprint,tightenlines,floatfix,showpacs,preprintnumbers,nofootinbib,eqsecnum,superscriptaddress]{revtex4}

 \usepackage[dvips,final]{graphicx}
  \usepackage{amssymb}
    \usepackage{amsfonts}
     \usepackage{epsfig}
      \usepackage{bm}

\def\Pom{{\bf I\!P}}
\newcommand{\bsigma}{\mbox{\boldmath $\sigma$}}

\newcommand{\bp}{\mbox{\boldmath $p$}}
\newcommand{\bq}{\mbox{\boldmath $q$}}

\newcommand{\bk}{\mbox{\boldmath $k$}}

\newcommand{\be}{\mbox{\boldmath $e$}}
\newcommand{\bn}{\mbox{\boldmath $n$}}
\newcommand{\bM}{\mbox{\boldmath $M$}}

\newcommand{\ket}[1]{| {#1} \rangle}
\newcommand{\bra}[1]{\langle {#1} |}

\newcommand{\half}{{1\over 2}}
\def\lsim{\mathrel{\rlap{\lower4pt\hbox{\hskip1pt$\sim$}}
    \raise1pt\hbox{$<$}}}         
\def\gsim{\mathrel{\rlap{\lower4pt\hbox{\hskip1pt$\sim$}}
    \raise1pt\hbox{$>$}}}         

\begin{document}

\thispagestyle{empty} \preprint{\hbox{}} \vspace*{-10mm}

\title{Exclusive photoproduction of charmonia \\
in $\gamma p \to V p$ and $p p \to p V p$ reactions \\
within $k_t$-factorization approach}

\author{A. Cisek}
\email{acisek@univ.rzeszow.pl}
\affiliation{University of Rzesz\'ow, PL-35-959 Rzesz\'ow,
Poland} 
\author{W. Sch\"afer}
\email{Wolfgang.Schafer@ifj.edu.pl}
\affiliation{Institute of Nuclear Physics PAN, PL-31-342 Cracow,
Poland} 
\author{A. Szczurek}
\email{Antoni.Szczurek@ifj.edu.pl}
\affiliation{Institute of Nuclear Physics PAN, PL-31-342 Cracow,
Poland} 
\affiliation{University of Rzesz\'ow, PL-35-959 Rzesz\'ow,
Poland}

\date{\today}

\begin{abstract}
The amplitude for $\gamma p \to J/\psi p$ ($\gamma p \to \psi' p$) 
is calculated in a pQCD $k_{T}$-factorization approach. 
The total cross section for this process is calculated for different
unintegrated gluon distributions and compared with the HERA data
and the data extracted recently by the LHCb collaboration.
The amplitude for $\gamma p \to J/\psi p$ ($\gamma p \to \psi' p$) 
is used to predict
the cross section for exclusive photoproduction of $J/\psi$ ($\psi'$) meson
in proton-proton collisions. Compared to earlier calculations we include
both Dirac and Pauli electromagnetic form factors.
The effect of Pauli form factor is quantified.
Absorption effects are taken into account and their role is discussed
in detail.
Different differential distributions e.g. in $J/\psi$ ($\psi'$)
rapidity and transverse momentum are presented and compared with existing
experimental data.
The UGDF with nonlinear effects built in better describe recent
experimental data of the LHCb collaboration but no definite conclusion
on onset of saturation can be drawn.
We present our results also for the Tevatron. A good agreement with the CDF
experimental data points at the midrapidity for both $J/\psi$ and $\psi'$
is achieved.
\end{abstract}


\maketitle

\section{Introduction}

The exclusive production of $J/\psi$
mesons in proton-proton and proton-antiproton scattering has recently
attracted some interest \cite{CDF,LHCb_first,LHCb_second,Klein:2003vd,SS2007,BMSC2007,MW2008,Goncalves:2011vf,
Ducati:2013tva,JMRT2013}. 

In an early paper \cite{SS2007}, it was shown that the 
exclusive production of $J/\psi$ at the Tevatron
is sensitive to the $\gamma p \to J/\psi p$ scattering in the similar
region of energy as measured at HERA \cite{HERA_new}. Given that fact
the measured cross section for $\gamma p \to J/\psi p$ was parametrized
and used in that calculation. Our predictions there could be successfully
confronted with the Tevatron data \cite{CDF} and good agreement was
achieved \cite{Cisek_phd}. The formalism proposed in \cite{INS} and 
used in \cite{SS2007} allows to calculate fully differential distributions
for the three-body reaction in the broad range of four-dimensional 
phase space.
The formalism proposed in \cite{IN} allows to test unintegrated gluon 
distributions (UGDFs) in the proton provided the quark-antiquark wave 
function of the meson is known. The experimental data for production
of different vector mesons prefer Gaussian light-cone wave function
\cite{INS,Ivanov:2003iy,RSS,CSS}.

Recently also the LHCb collaboration measured rapidity distributions
of the $J/\psi$ meson but rather in semi-exclusive reaction
\cite{LHCb_first,LHCb_second} \footnote{The protons were not
detected and only incomplete rapidity gap was checked.}. 
Using some theoretical input from Ref.\cite{SS2007} the LHCb collaboration
tried to extract the cross for the $\gamma p \to J/\psi p$ reaction
at unprecedently high energies not available before at HERA,
The procedure proposed uses some assumption which are only approximate
and need further verification.
Very recently the authors of Ref.\cite{JMRT2013} tried to use the
pseudo-data to achieve information on integrated gluon distribution
in very small $x$ (longitudinal momentum fraction carried by the gluon) 
region, not available earlier at electron machines.
The formalism applied in Ref.\cite{JMRT2013} uses a slightly simplified 
collinear formalism where the quark-antiquark wave function is
replaced by a normalization constant \cite{Ryskin}. In this formalism 
only rapidity distribution was discussed. 
In contrast to the collinear approach the $k_t$-factorization approach
allows to study the complete kinematically reaction. In the present
analysis we shall show how some UGDFs from the literature compare to 
the LHCb data \cite{LHCb_first,LHCb_second}. We leave the inclusion of the inelastic contribution
as well as a possible fitting of UGDF for further sudies.
Compared to other calculations in the literature we include here
not only the spin-conserving coupling but also the spin-flip one.

\section{Photoproduction process $\gamma p \to J/\psi p$}

\begin{figure}[!htb] 
\begin{center}
\includegraphics[height=6.0cm]{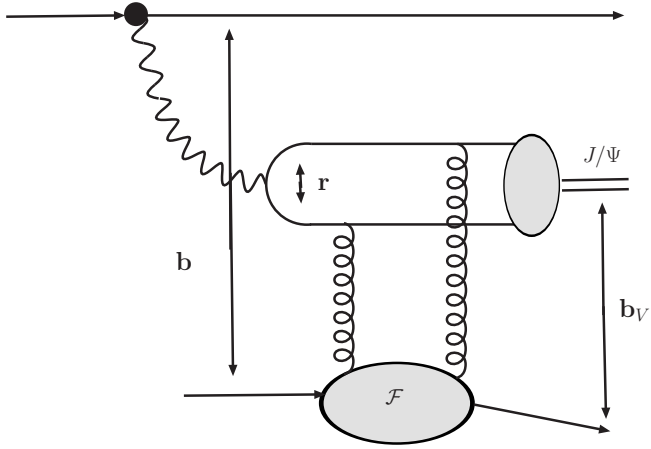}
\caption[*]{
Diagrams representing amplitude for the $\gamma p \to  J/\psi$ p process.
\label{fig:gammap_Vp}
}
\end{center}
\end{figure}

The imaginary part of the forward amplitude sketched in Fig.\ref{fig:gammap_Vp} 
can be written as \cite{Ivanov:2003iy,INS}:

\begin{eqnarray}
\Im m \, {\cal M}_{T}(W,\Delta^2 = 0,Q^{2}=0) =
W^2 \frac{c_v \sqrt{4 \pi \alpha_{em}}}{4 \pi^2} \, 2 \, 
 \int_0^1 \frac{dz}{z(1-z)}
\int_0^\infty \pi dk^2 \psi_V(z,k^2)
\nonumber \\
\int_0^\infty
 {\pi d\kappa^2 \over \kappa^4} \alpha_S(q^2) {\cal{F}}(x_{\rm eff},\kappa^2)
\Big( A_0(z,k^2) \; W_0(k^2,\kappa^2) 
     + A_1(z,k^2) \; W_1(k^2,\kappa^2)
\Big) \, ,
\end{eqnarray}

where, for the pure $S$-wave vector meson,

\begin{eqnarray}
A_0(z,k^2) &=& m_c^2 + \frac{k^2 m_q}{M + 2 m_c}  \, ,
\nonumber \\
A_1(z,k^2) &=& \Big[ z^2 + (1-z)^2 
    - (2z-1)^2 \frac{m_c}{M + 2 m_c} \Big] \, \frac{k^2}{k^2+ m_c^{2}} \, ,
\nonumber
\end{eqnarray}

\begin{eqnarray}
W_0(k^2,\kappa^2) &=& 
{1 \over k^2 + m_c^2} - {1 \over \sqrt{(k^2-m_c^2-\kappa^2)^2 + 4 m_c^2 k^2}}
\, , 
\nonumber \\
W_1(k^2,\kappa^2) &=& 1 - { k^2 + m_c^2 \over 2 k^2}
\Big( 1 + {k^2 - m_c^2 - \kappa^2 \over 
\sqrt{(k^2 - m_c^2 - \kappa^2)^2 + 4 m_c^2 k^2 }}
\Big) \, .
\nonumber
\end{eqnarray}

Here $\psi_V(z,k^2)$ is the meson light-cone wave function,
${\cal{F}}(x_{\rm eff},\kappa^{2})$ is the unintegrated gluon distribution function.
The invariant mass of the $c \bar c$-system is given by
\begin{eqnarray}
M = \sqrt{ {k^2 + m_c^2 \over z (1-z)}} 
\end{eqnarray}

We choose the scale of the QCD constant running coupling at
$q^2 = \max \{ \kappa^2, k^2 + m_c^2 \}$.

The full amplitude, at finite momentum transfer is given by:
\begin{eqnarray}
{\cal M}(W,\Delta^2) = (i + \rho) \, \Im m {\cal M}(W,\Delta^2=0,Q^{2}=0)
\, \exp(-B(W) \Delta^2/2) \, ,
\label{full_amp}
\end{eqnarray}
where the real part of the amplitude is restored from analyticity,
\begin{eqnarray}
\rho = {\Re e {\cal M} \over \Im m {\cal M}} =  
\tan \Big ( {\pi \over 2} \, { \partial \log \Big( \Im m {\cal M}/W^2 \Big) \over \partial \log W^2 } \Big) \, .
\end{eqnarray}
Above $B(W)$ is a slope parameter which in general depends on the photon-proton
center-of-mass energy and is parametrized in the present analysis as:
\begin{eqnarray}
B(W) = b_0 + 2 \alpha'_{eff} \log \Big( {W^2 \over W^2_0} \Big) \, ,
\end{eqnarray}
with: $b_0 = 4.88$, $\alpha'_{eff} = 0.164$ GeV$^{-2}$ and $W_0 = 90$
GeV \cite{HERA_new}.

Finally, the total cross section for diffractive $J/\psi$ photoproduction on the nucleon 
can be calculated from:
\begin{eqnarray}
\sigma(\gamma p \to J/\psi p) = {1 +  \rho^2 \over 16 \pi B(W)} 
\Big| \Im m { {\cal M}(W,\Delta^2=0) \over W^2 } \Big|^2 \, .
\end{eqnarray}
%

\section{$J/\psi$ and $\psi'$ wave functions}

We include the Fermi-motion of quark and antiquark in the bound-state by means of the 
light-cone wave function of the vector meson. 

While the light-cone wave function $ \psi_V(z,k^2)$  is written as 
a function of the momentum fraction $z$ of the quark and the relative transverse momentum $\bk$
of quarks in the bound state, in fact it depends only on the relative momentum $\vec{p}$
of $c$ and $\bar c$ in the rest frame of the meson given by
\begin{eqnarray}
 \vec{p} = (\bp , p_z) = (\bk, (z-1/2) M) \; .
\end{eqnarray}
Following quite literally the
approach of \cite{Ivanov:2003iy,INS} we use as an ansatz for the wave function dependence on $\vec{p}$:
\begin{eqnarray}
 \psi_{\rm 1S} (z,k^2) &=& \psi_{\rm 1S} (\vec{p}\,^2) = c_1 \,
 \exp\Big( - a_1^2 \vec{p}\,^2 /2 \Big) \; , \nonumber \\
 \psi_{\rm 2S} (z,k^2) &=& \psi_{\rm 2S} (\vec{p}\,^2) = c_2 (\xi_{\rm node} - a_2^2 \vec{p}\,^2)  \, \exp\Big( - a_2^2 \vec{p}\,^2 /2 \Big) \, .\nonumber \\
\end{eqnarray}
This functional dependence is obviously inspired by the harmonic-oscillator potential, notice however
that following \cite{Ivanov:2003iy,INS} we keep $a_1,a_2$ which would be equal in the strict harmonic oscillator
potential as free parameters. The parameters are fixed from the leptonic decay widths of $J/\psi$ and $\psi'$,
as well as from the orthonormality conditions of the $1S,2S$ bound states.
Recently a number of theoretical arguments for an effectively harmonic confinement potential
on the light-front from various approaches have been given in \cite{Trawinski:2014msa}. 

Let us stress that an account of the wave function is important to make 
predictions for the production of excited charmonium states. While we use 
the momentum space formulation of diffractive vector meson production, 
the equivalent color-dipole formulation is more intuitive to
understand the argument: it is the overlap of the light-cone wave
functions of photon and vector meson which controls the effective dipole
size distributions that enter the dipole cross section 
\cite{Nikolaev:1992si,Kopeliovich:1993gk}. Here especially the node 
in the wave function of the radial excitation has a subtle effect on 
the energy dependence \cite{Nemchik:1996cw}.

Often an extreme nonrelativistic limit is adopted, in which heavy quarks are assumed to be at rest
in the meson rest frame, and hence the momentum dependence of the wave function is neglected.
Typically then also the transverse momentum of gluons is integrated out \cite{Ryskin}, and
the diffractive amplitude becomes proportional to the integrated glue of the target.
Strictly speaking in such an approximation one cannot predict the energy dependence
of $J/\psi$ vs. $\psi'$ production, as to the accuracy of \cite{Ryskin} it is illegitimate to differentiate between
$2 m_c$, the invariant mass of the $c \bar c$ pair, $M$, or the bound-state mass $M_V$, either
of which could enter the hard scale. The wave function ``node effect'' \cite{Kopeliovich:1993gk,Ivanov:2003iy},
which leads to a strong dependence of the $\psi'/J/\psi$ ratio on the bound-state wave function
clearly cannot be accomodated in this way.

\section{Exclusive photoproduction of $J/\psi$ in $p p$ and $p \bar p$ collisions}
\begin{figure}[!htb] 
\begin{center}
\includegraphics[height=5.0cm]{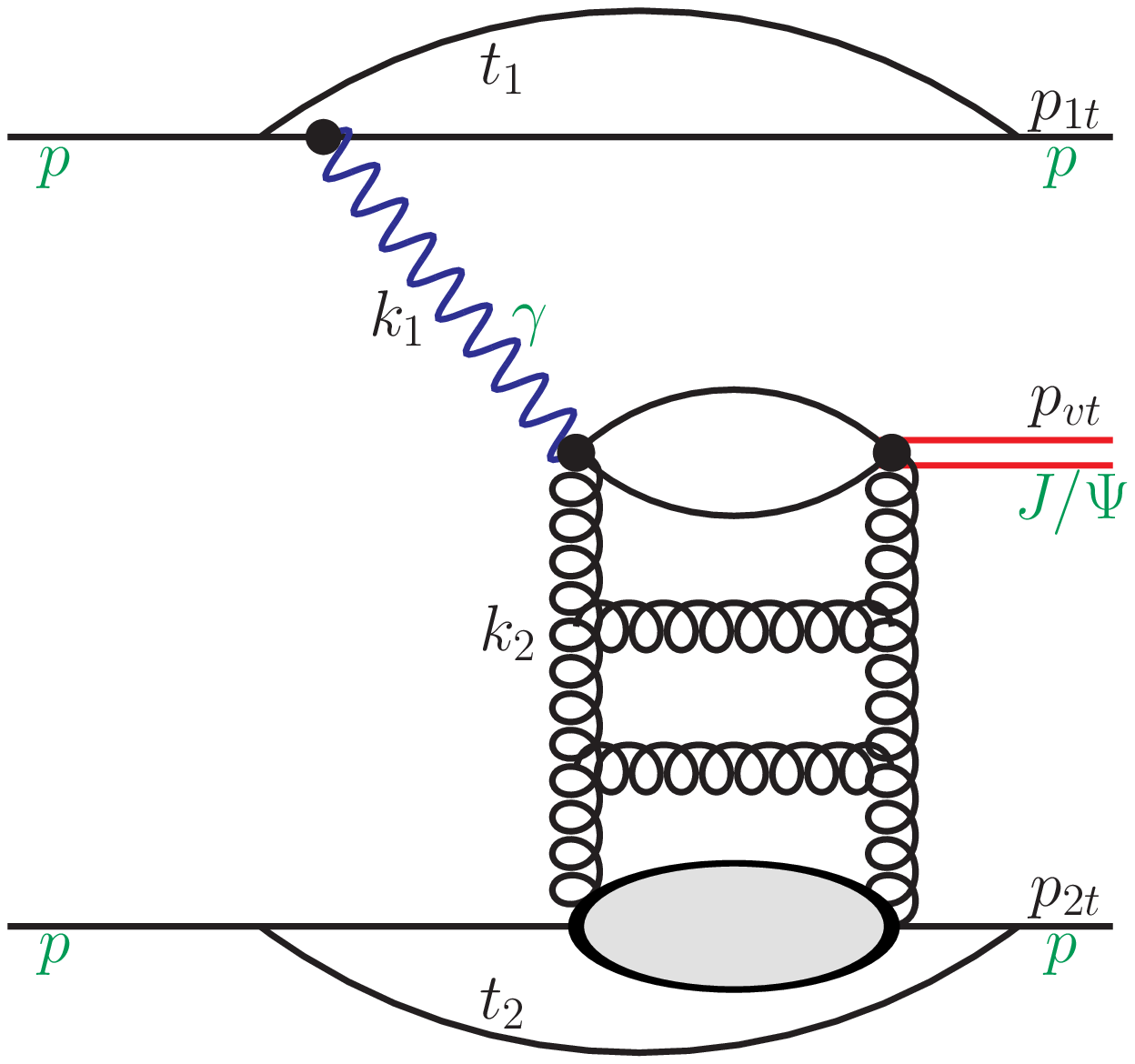}
\includegraphics[height=5.0cm]{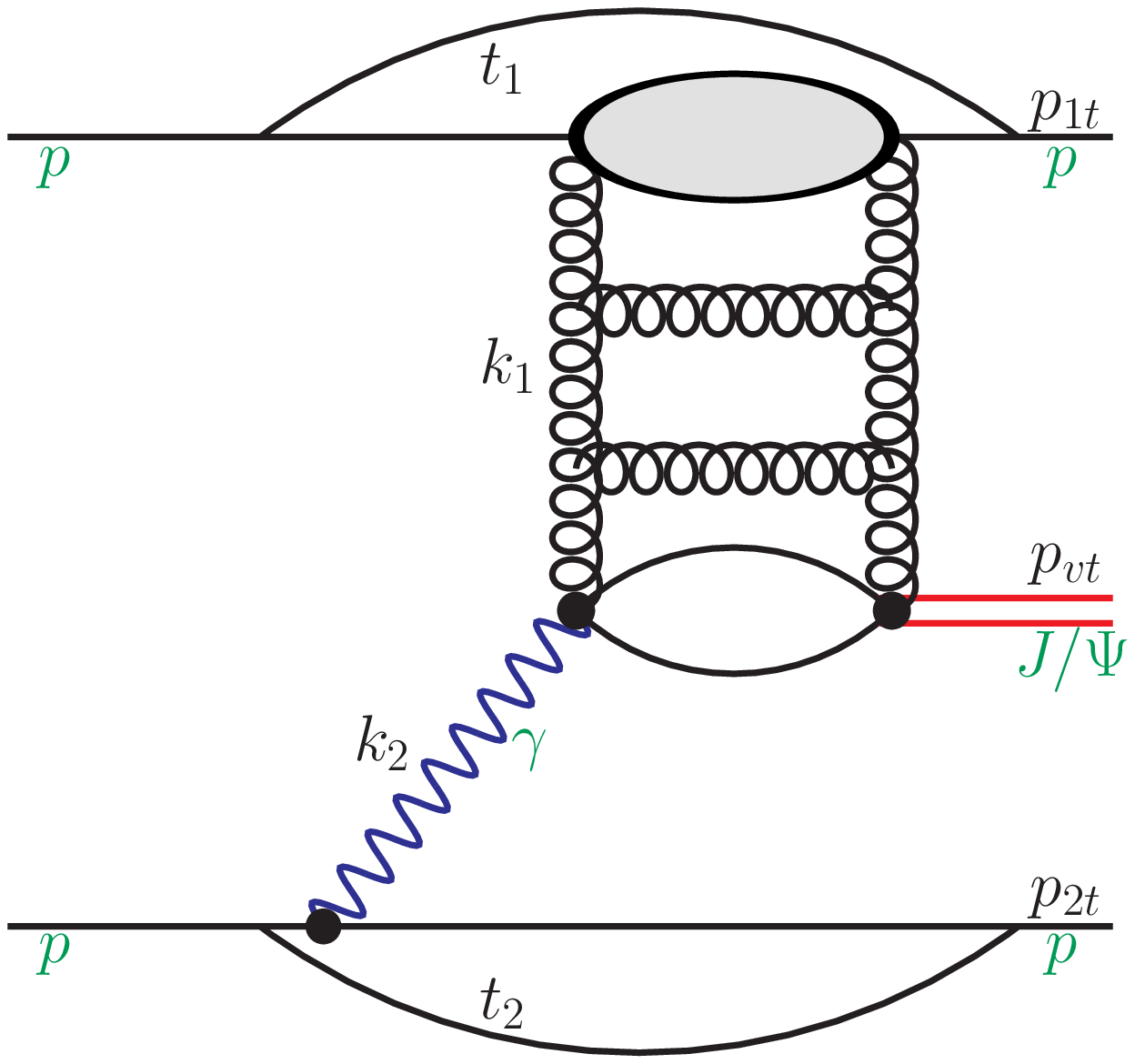}
\caption[*]{
Diagrams representing Born amplitudes considered
for the $p p \to p p J/\psi$ process.
\label{fig:diagrams_Born}
}
\end{center}
\end{figure}

The Born mechanism of the production of the $J/\psi$ meson
(similar mechanism for $\psi'$) in proton-proton collisions is shown in
Fig.\ref{fig:diagrams_Born}. 
There are two diagrams. In first diagram photon couples to the first
proton while in the second diagram the photon couples to the second
proton. The photon splits to a $c \bar c$ dipole which interacts
with the other proton via exchange of gluonic ladder.
The presence of two mechanisms leads to interference effects \cite{SS2007}.
The interference effect leads to interesting azimuthal correlations
between outgoing protons \cite{SS2007}, never identified so far.

The full Born amplitude for the $ pp \to pVp$ process can be written as:
\begin{eqnarray}
{\cal M}_{h_1 h_2 \to h_1 h_2 V}^
{\lambda_1 \lambda_2 \to \lambda'_1 \lambda'_2 \lambda_V}(s,s_1,s_2,t_1,t_2) =
{\cal M}_{\gamma \Pom} + {\cal M}_{\Pom \gamma} \nonumber \\
= \bra{p_1', \lambda_1'} J_\mu \ket{p_1, \lambda_1} 
\epsilon_{\mu}^*(q_1,\lambda_V) {\sqrt{ 4 \pi \alpha_{em}} \over t_1}
{\cal M}_{\gamma^* h_2 \to V h_2}^{\lambda_{\gamma^*} \lambda_2 \to \lambda_V \lambda_2}
(s_2,t_2,Q_1^2)   \nonumber \\
 + \bra{p_2', \lambda_2'} J_\mu \ket{p_2, \lambda_2} 
\epsilon_{\mu}^*(q_2,\lambda_V)  {\sqrt{ 4 \pi \alpha_{em}} \over t_2}
{\cal M}_{\gamma^* h_1 \to V h_1}^{\lambda_{\gamma^*} \lambda_1 \to \lambda_V \lambda_1}
(s_1,t_1,Q_2^2)  \, .
\label{Two_to_Three}
\end{eqnarray}

In terms of their transverse momenta $\bp_{1,2}$ the relevant
four--momentum transfers squared are $t_1 = -(\bp_1^2 + z_1^2 m_p^2)/(1-z_1)$
and $t_2 = -(\bp_2^2 + z_2^2 m_p^2)/(1-z_2)$ and $s_1 \approx (1 -z_2) s$ and
$s_2 \approx (1-z_1) s$ are the familiar Mandelstam variables.

Then, the amplitude of Eq. (\ref{Two_to_Three}) for the emission of a photon
of transverse polarization $\lambda_V$, and transverse momentum
$\bq_1 = - \bp_1$ can be written as:

\begin{eqnarray} 
\bra{p_1', \lambda_1'} J_\mu \ket{p_1, \lambda_1} 
\epsilon_{\mu}^*(q_1,\lambda_V)
={ (\be^{*(\lambda_V)} \bq_1)  \over \sqrt{1-z_1}} 
\, {2 \over z_1} \, \chi^\dagger_{\lambda'} 
\Big\{  F_1(Q_1^2) 
- {i \kappa_p F_2(Q_1^2) \over 2 m_p}  
( \bsigma_1 \cdot [\bq_1,\bn]) \Big\} \chi_\lambda \, .
\end{eqnarray}

Above $\chi_\lambda$ is its spinor,
$\be^{(\lambda)} = -(\lambda \be_x + i \be_y)/\sqrt{2}$, $\bn || \be_z$
denotes the collision axis, and $\bsigma_1/2$ is the spin operator for nucleon $1$.
$F_1$ and $F_2$ are respectively the Dirac and Pauli electromagnetic form factors,
$\kappa_p = 1.79$.

Below the $2 \to 3$ bare amplitude (when absorption effects is ignored)
is shown in the form of a 2--dimensional vector:
\begin{eqnarray}
\bM^{(0)}(\bp_1,\bp_2) &&= e_1 {2 \over z_1} {\bp_1 \over t_1} 
{\cal{F}}_{\lambda_1' \lambda_1}(\bp_1,t_1)
{\cal {M}}_{\gamma^* h_2 \to V h_2}(s_2,t_2,Q_1^2)
\nonumber \\
&&
+ e_2 {2 \over z_2} {\bp_2 \over t_2} {\cal{F}}_{\lambda_2' \lambda_2}(\bp_2,t_2)
{\cal {M}}_{\gamma^* h_1 \to V h_1}(s_1,t_1,Q_2^2)
\, .
\nonumber
\end{eqnarray}
Because of the presence of the proton from factors only small 
$Q_{1}^{2}$ and $Q_{2}^{2}$
enter the amplitude for the hadronic process. This means that in 
practice, inside the photoproduction amplitude, one can put
$Q_{1}^{2} = Q_{1}^{2} = 0 $.

\begin{figure}[!htb] 
\begin{center}
\includegraphics[height=5.0cm]{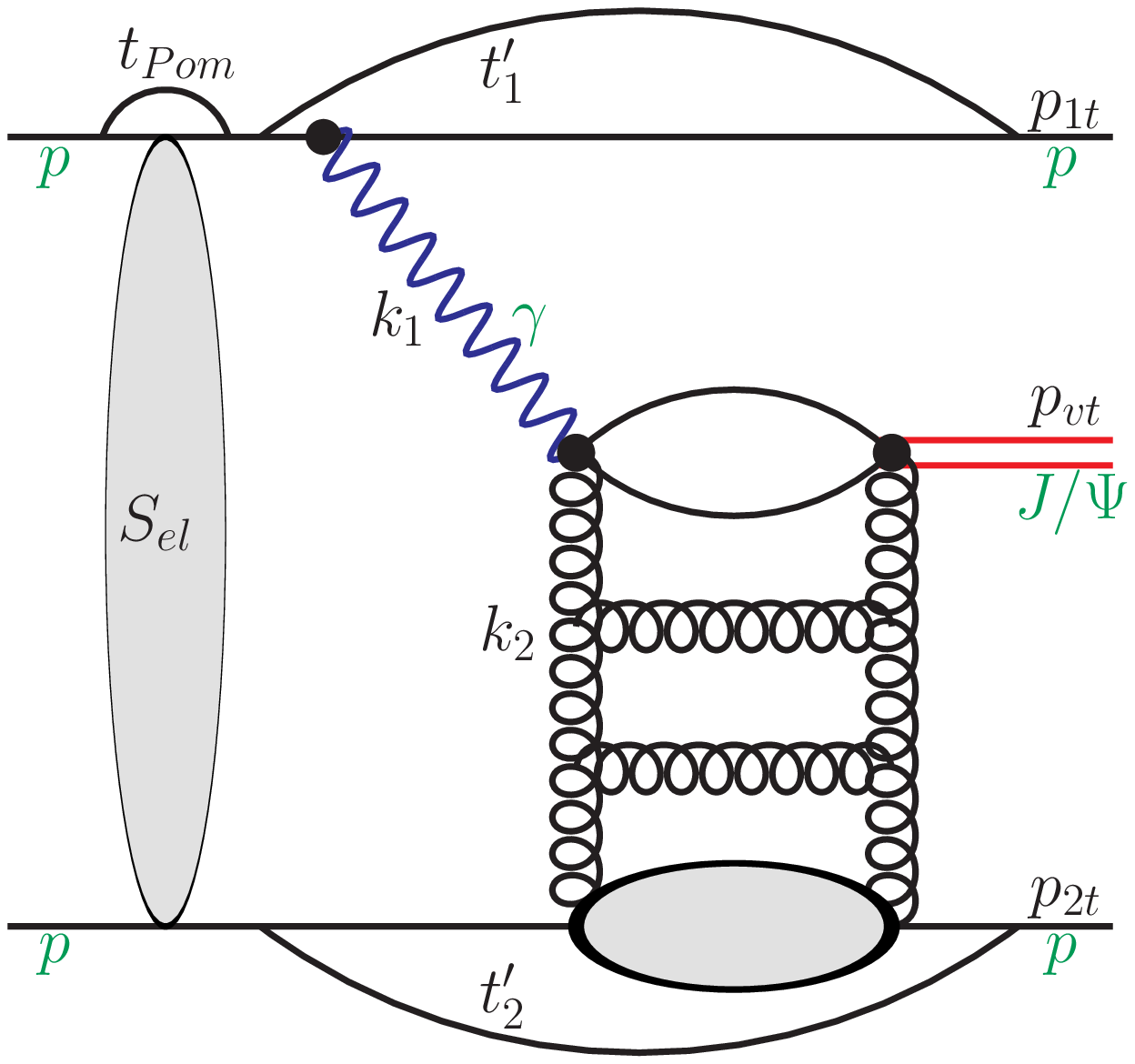}
\includegraphics[height=5.0cm]{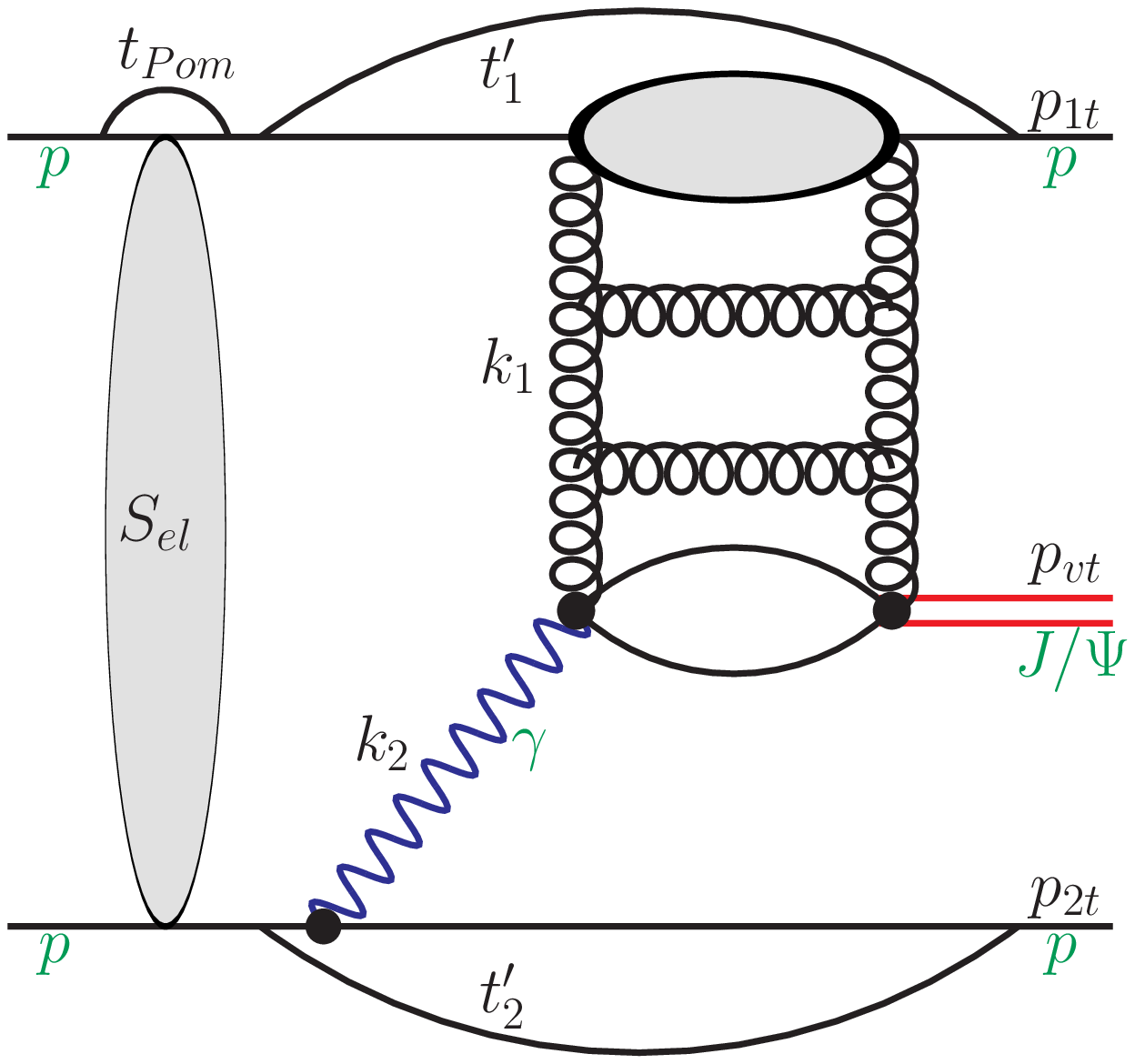}
\caption[*]{
Diagrams representing absorptive corrections considered
for the $p p \to p p J/\psi$ process.
\label{fig:diagrams_absorptive}
}
\end{center}
\end{figure}

The full amplitude for the $p p \to p J/\psi p$ 
or $p p to p \phi' p$ is calculated as:
\begin{eqnarray}
\bM(\bp_1,\bp_2) &&= \int{d^2 \bk \over (2 \pi)^2} \,S_{el}(\bk) \,
\bM^{(0)}(\bp_1 - \bk, \bp_2 + \bk)  \nonumber \\
&&= \bM^{(0)}(\bp_1,\bp_2) - \delta \bM(\bp_1,\bp_2)\, .
\end{eqnarray}
The corresponding diagrams are shown in Fig.\ref{fig:diagrams_absorptive}.
In the present calculations we incude only elastic rescattering
corrections.
Then
\begin{equation}
S_{el}(\bk) = (2 \pi)^2 \delta^{(2)}(\bk) - \half T(\bk) \, \, \, ,
\, \, \, T(\bk) = \sigma^{pp}_{tot}(s) \, \exp\Big(-\half B_{el} \bk^2 \Big) \, .
\nonumber
\end{equation}
In practical evaluations we take $B_{el} = 17$ GeV$^{-2}$, 
\ $\sigma^{pp}_{tot} = 76$ mb \cite{CDF94,SS2007} for the Tevatron energy and 
$B_{el} = 19.89$ GeV$^{-2}$, \ $\sigma^{pp}_{tot} = 98.6$ mb 
for the LHC energy \cite{TOTEM}.

The absorptive correction to the amplitude can be written as:
\begin{eqnarray}
\delta \bM(\bp_1,\bp_2) = \int {d^2\bk \over 2 (2\pi)^2} \, T(\bk) \,
\bM^{(0)}(\bp_1-\bk,\bp_2+\bk) \, .
\nonumber
\end{eqnarray}

Inelastic intermediate proton excitations can be taken into account
effectively by multiplying elastic amplitdes by a constant bigger than 1.

\section{Results}

\subsection{$J/\psi$ production}

\begin{figure}[!htb] 
\begin{center}
\includegraphics[height=6.0cm]{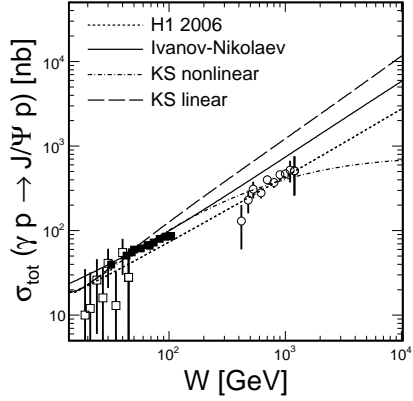}
\caption[*]{ 
Total cross section for the $\gamma p \to J/\psi p$ as a function of
the subsystem energy together with the HERA data and pseudodata
obtained by the LHCb collaboration. Three disfferent UGDFs have been
used: Ivanov-Nikolaev (solid), Kutak-Stasto linear (dashed) and 
Kutak-Stasto nonlinear (dash-dotted). The dotted line represents
calculation with a simple power-like parametrization of the old HERA
data \cite{H1_b}.
The HERA data points \cite{HERA_new} and the LHCb data ponits \cite{LHCb_first,LHCb_second}
are shown for comparison.
\label{fig:sig_tot_W}
}
\end{center}
\end{figure}

Before we go to the proton-proton processes let us first summarize
our description of HERA data \cite{H1_a,H1_b,HERA_new}.
In Fig.\ref{fig:sig_tot_W} we show results of calculations with
the Ivanov-Nikolaev \cite{IN,INS} and Kutak-Sta\'sto \cite{KS} gluon
unintegrated distributions. In the second case we consider both
a BFKL version called linear and a version where gluon is obtained by solving
Balitsky-Kovchegov evolution equation called nonlinear.

\begin{figure}[!htb] 
\begin{center}
\includegraphics[width=5.0cm]{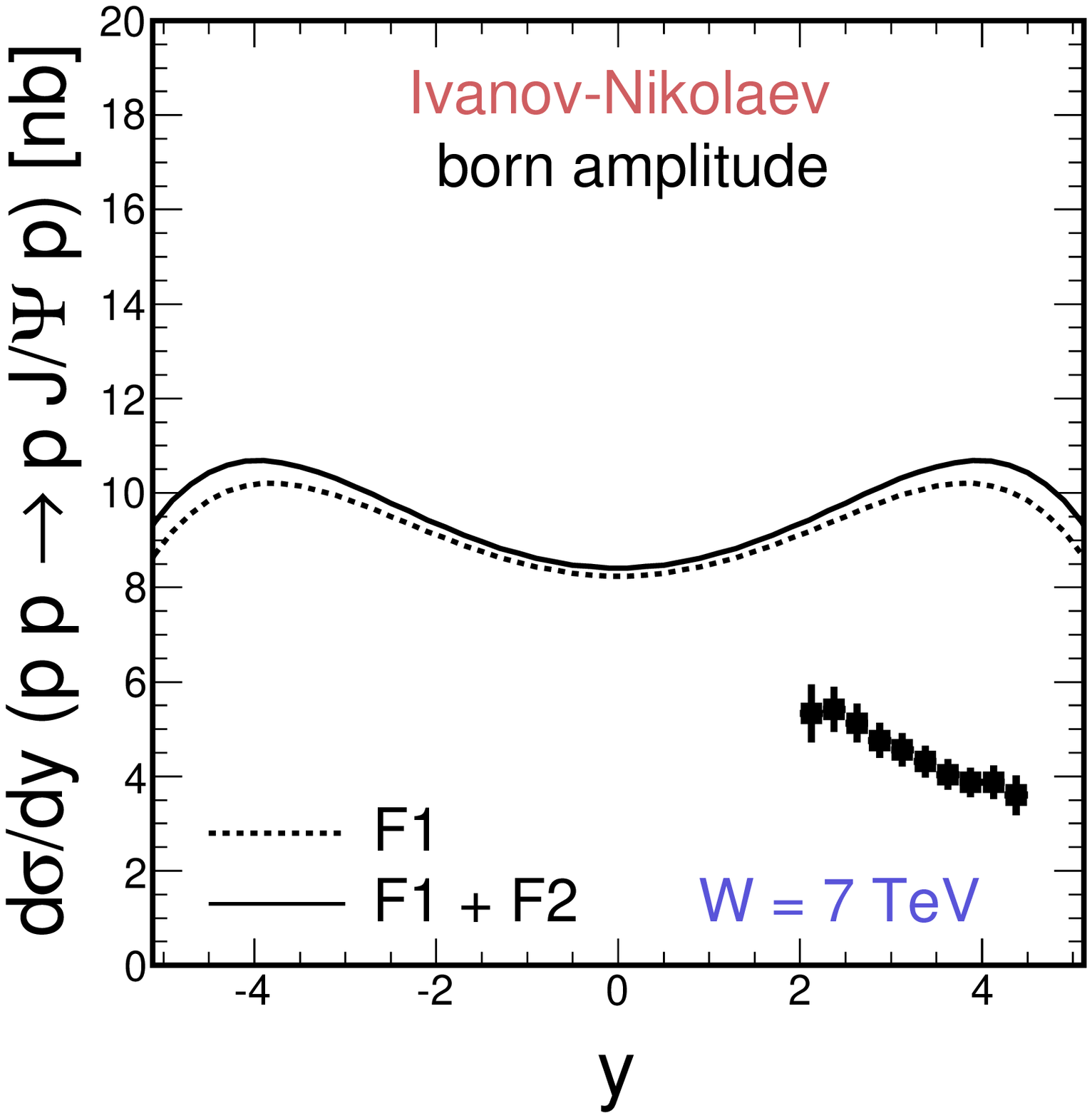}
\includegraphics[width=5.0cm]{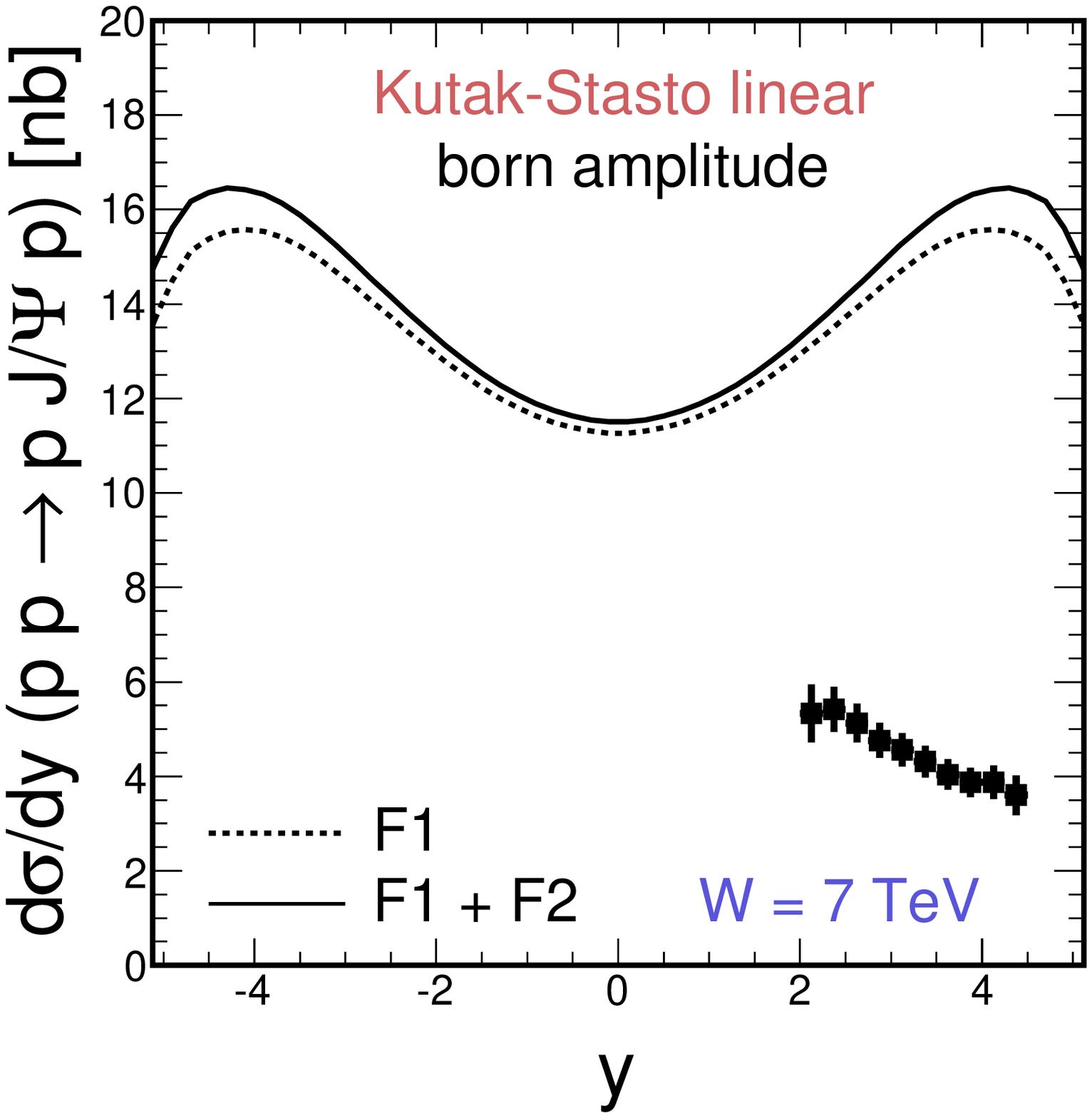}
\includegraphics[width=5.0cm]{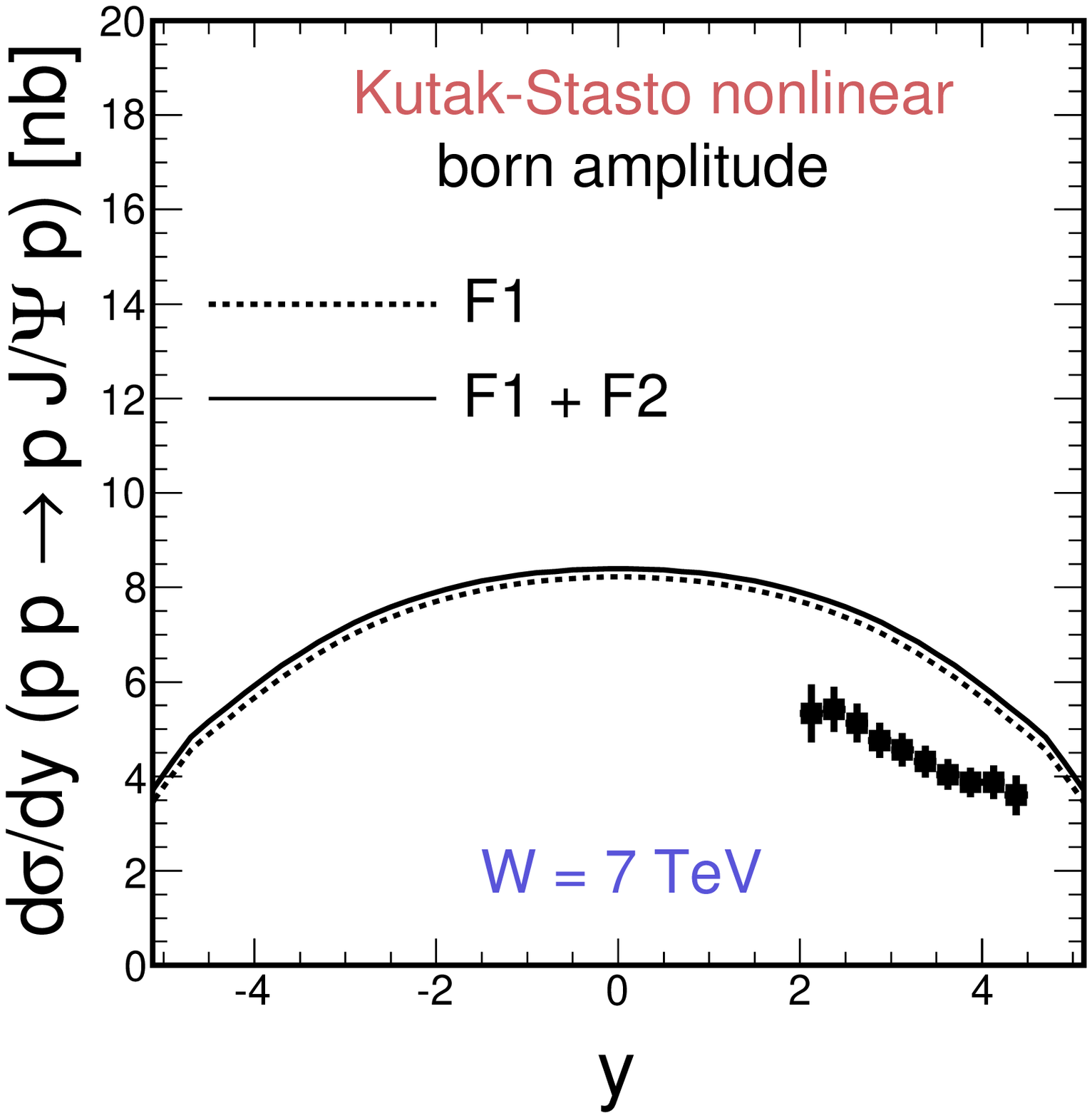}

\caption[*]{ 
$J/\psi$ rapidity distribution calculated with the Born amplitudes
for three different UGDFs from the literature for $\sqrt{s}$ = 7 TeV.
The dashed lines include contributions with Dirac $F_1$ electromagnetic 
form factor and the solid lines include in addition Pauli $F_2$ 
electromagnetic form factor.
The new LHCb data points \cite{LHCb_second} are shown for comparison.
\label{fig:dsig_dy_Born}
}
\end{center}
\end{figure}

In Fig.\ref{fig:dsig_dy_Born} we show rapidity distribution in the Born
approximation for different UGDFs. The dashed line represents calculation
when only vector ($F_1$) terms are included, while the solid line represents
calculations with vector ($F_1$) and tensor ($F_2$) couplings of photon
to the proton. The effect of taking into account tensor coupling is here
of the order of 5 \% only.

\begin{figure}[!htb] 
\begin{center}
\includegraphics[width=5.0cm]{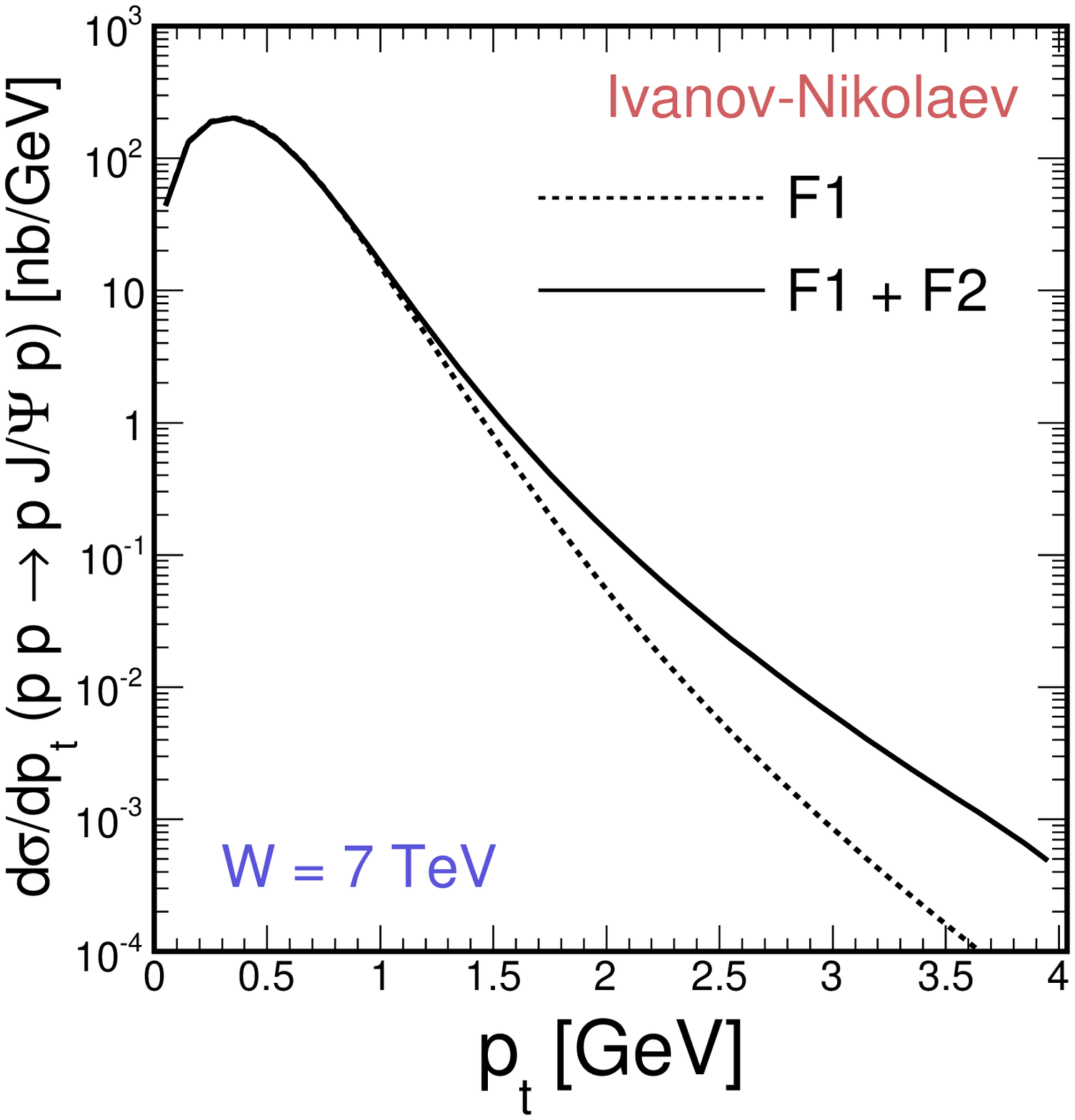}
\includegraphics[width=5.0cm]{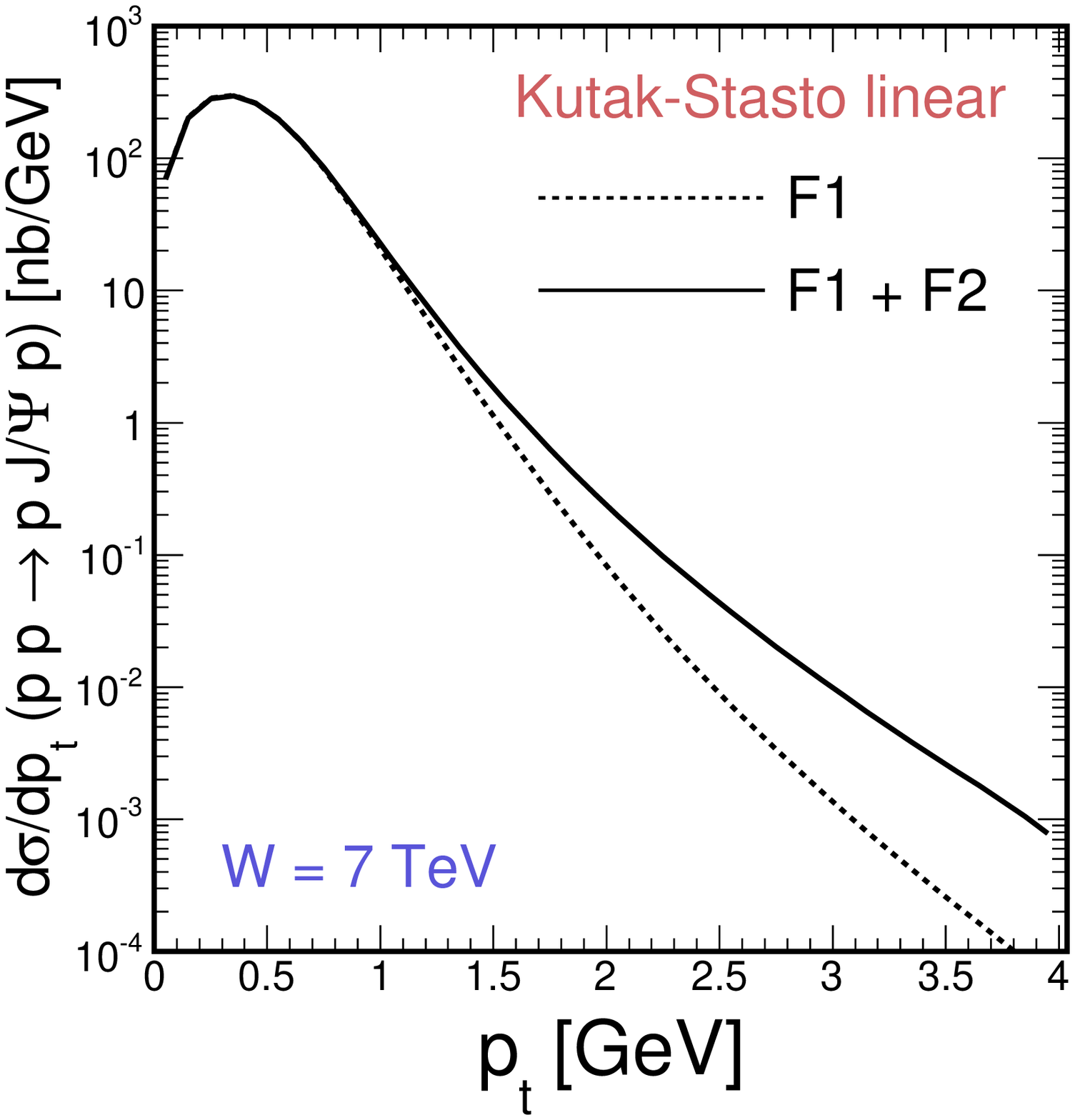}
\includegraphics[width=5.0cm]{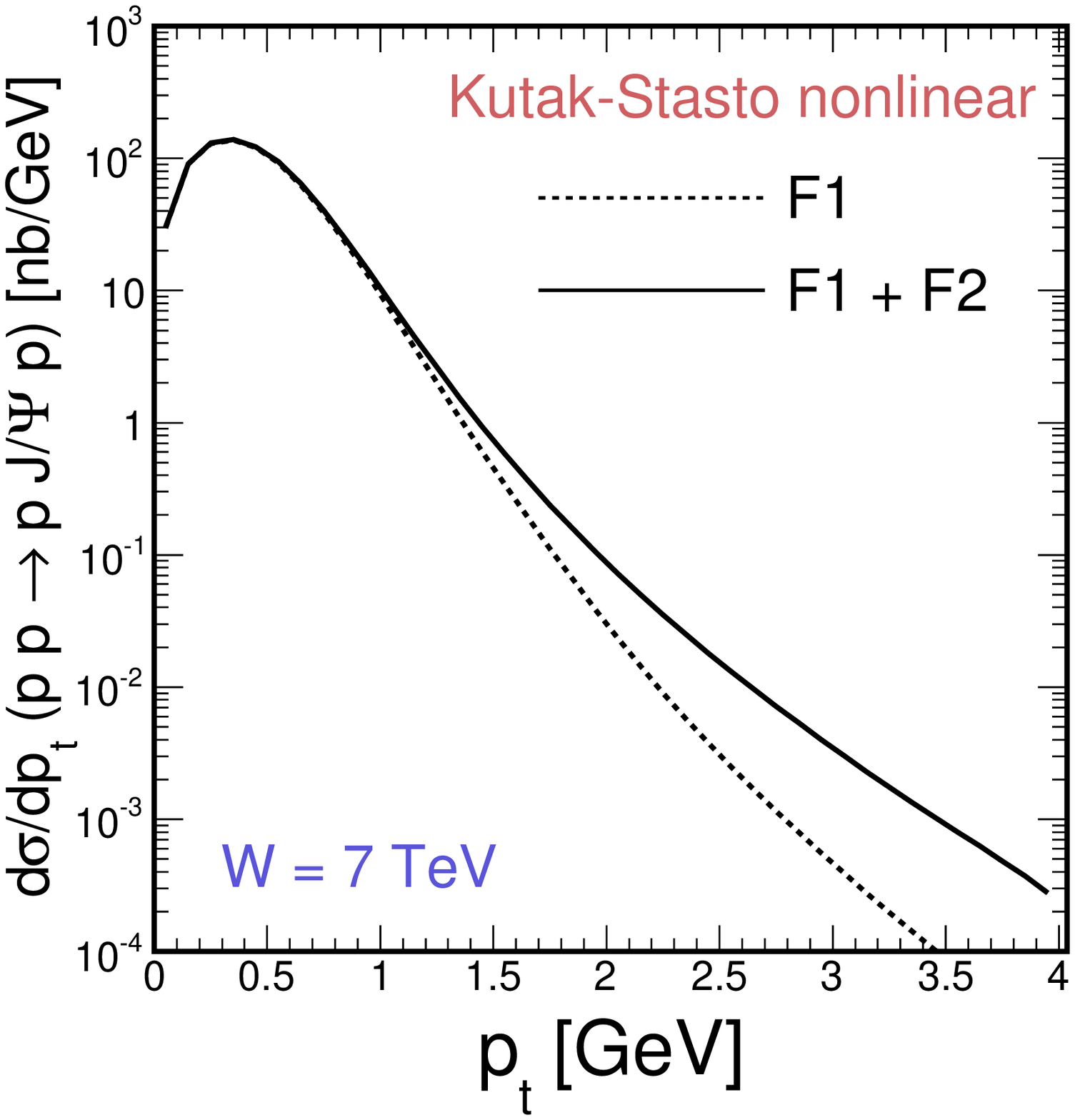}
\caption[*]{ 
$J/\psi$ transverse momentum distribution calculated with 
the Born amplitudes for three different UGDFs from the literature
for $\sqrt{s}$ = 7 TeV.
The dashed lines include contributions with Dirac $F_1$ electromagnetic 
form factor and the solid lines include in addition Pauli $F_2$ 
electromagnetic form factor.
\label{fig:dsig_dpt_Born}
}
\end{center}
\end{figure}

Similar distributions in $J/\psi$ transverse momentum are shown in
Fig.\ref{fig:dsig_dpt_Born}. Large effect of the tensor coupling can
be observed at large transverse momenta. At $p_t \sim$ GeV we get an
enhancement factor of the cross section of order of 10.
Large transverse momenta are potentially interesting because
of odderon exchange contribution (see e.g.\cite{BMSC2007}).

\begin{figure}[!htb] 
\begin{center}
\includegraphics[width=5.0cm]{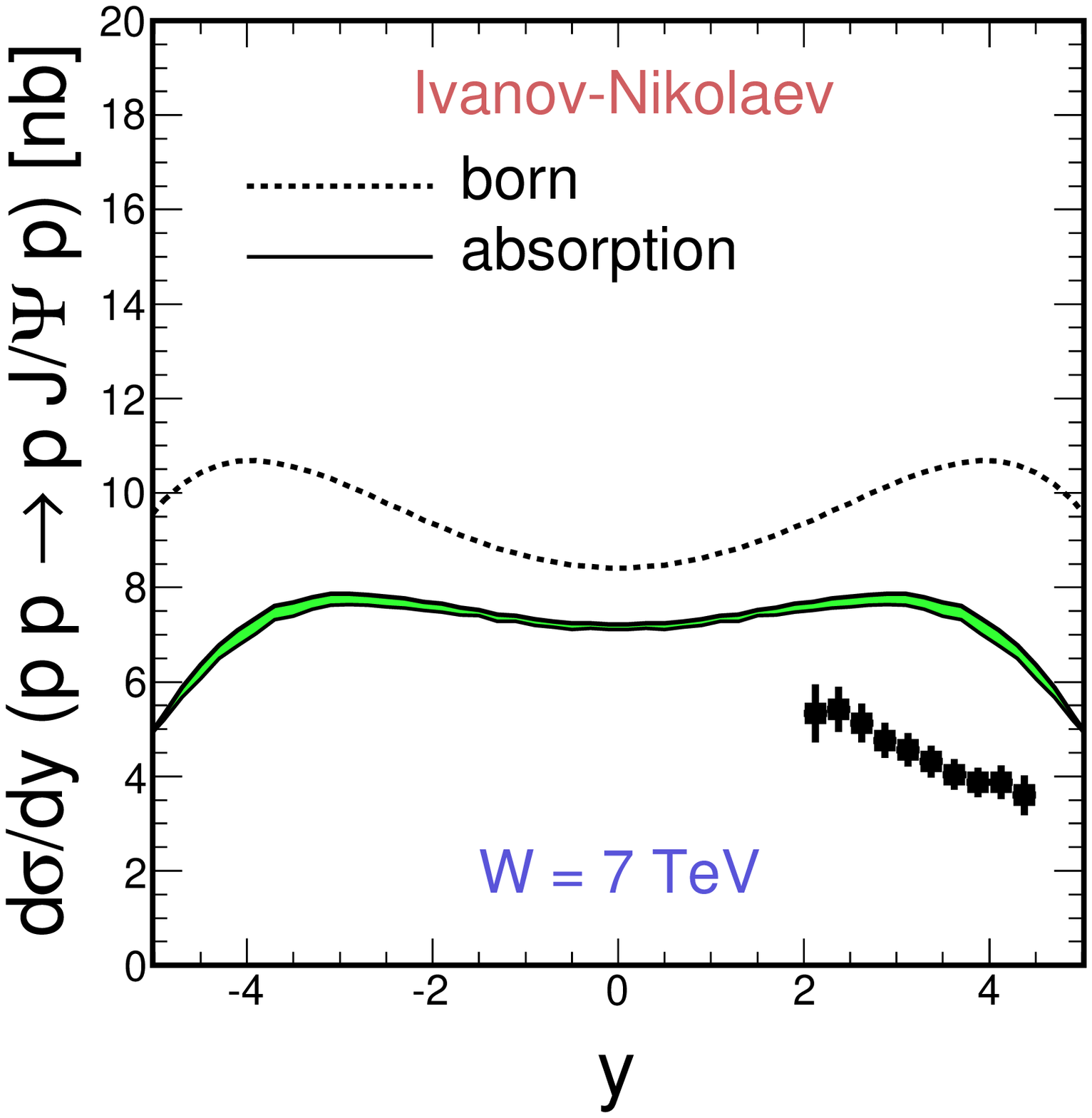}
\includegraphics[width=5.0cm]{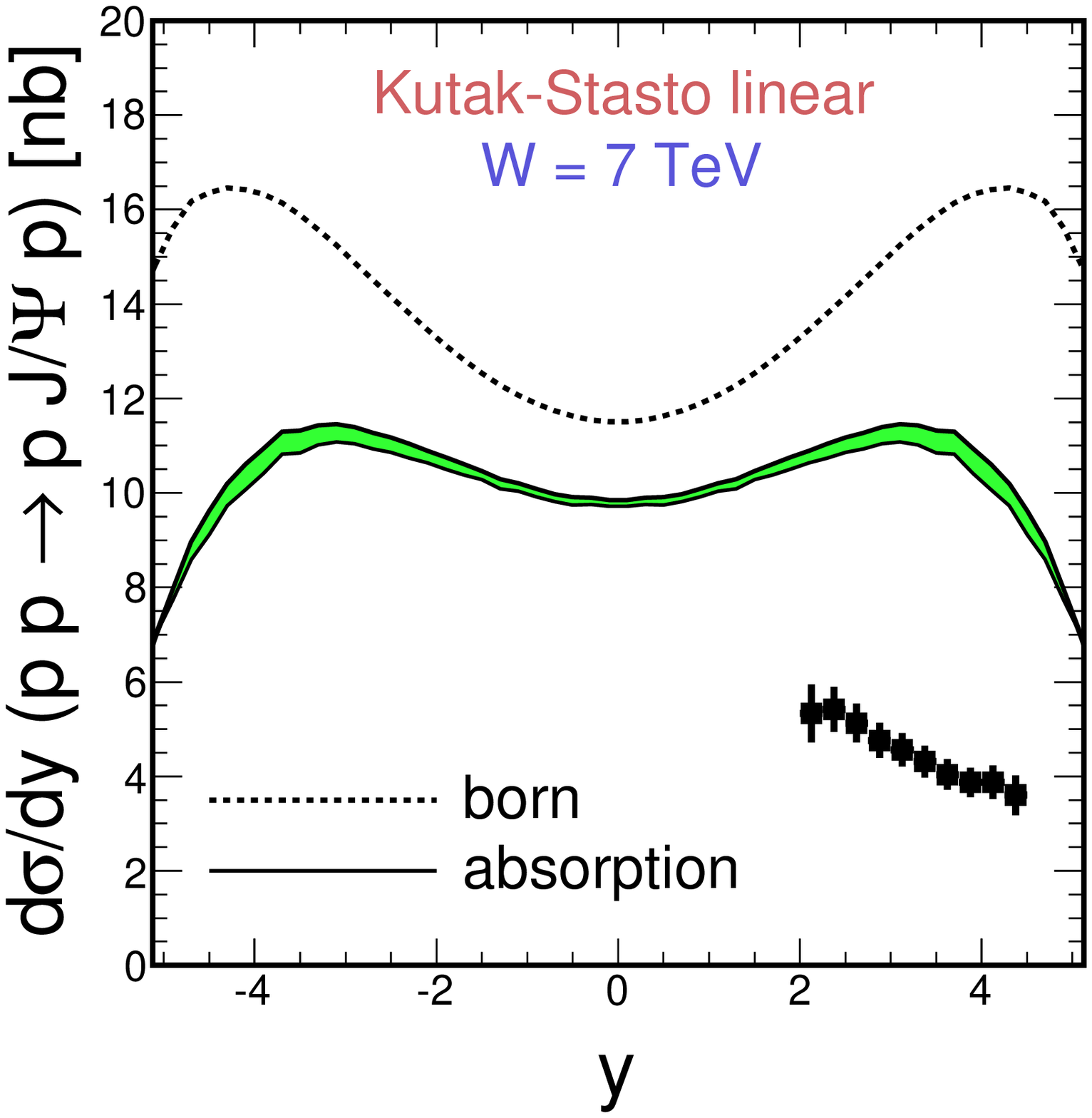}
\includegraphics[width=5.0cm]{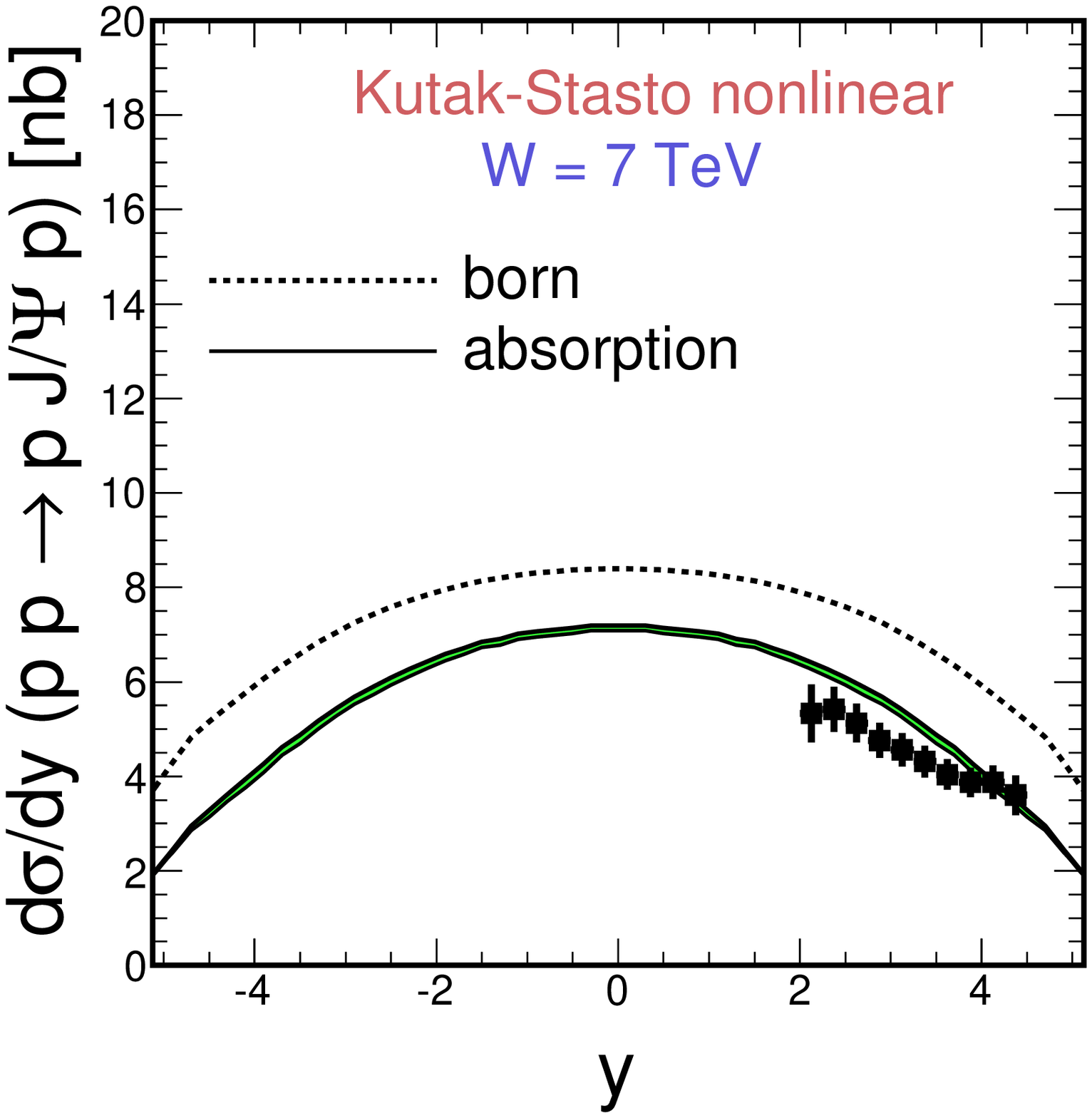}

\caption[*]{ 
$J/\psi$ rapidity distribution calculated with inclusion of 
absorption effects (solid line), compared with the Born result 
(dashed line) for $\sqrt{s}$ = 7 TeV. 
The new LHCb data points \cite{LHCb_second} are shown for comparison.
\label{fig:dsig_dy_absorption}
}
\end{center}
\end{figure}

The eikonal absorption damps rapidity distribution of $J/\psi$ by
about 30 \% as is shown in Fig.\ref{fig:dsig_dy_absorption}.
The result with the Kutak-Sta\'sto distribution which includes nonlinear
effects is almost consistent with the newest LHCb data.
Does it mean that we observe an onset of gluon saturation?
We shall return to it in a while.

\begin{figure}[!htb] 
\begin{center}
\includegraphics[width=5.0cm]{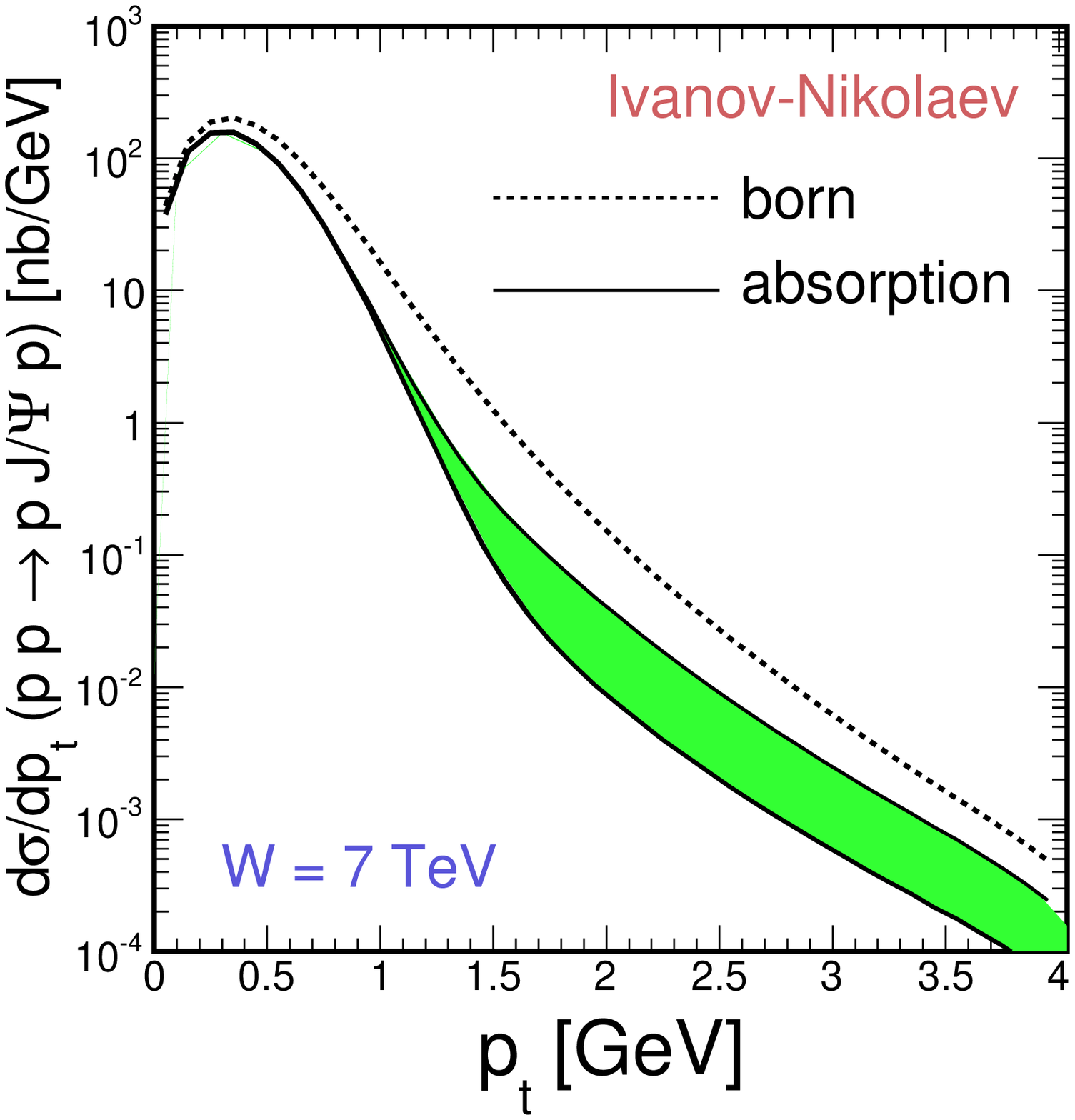}
\includegraphics[width=5.0cm]{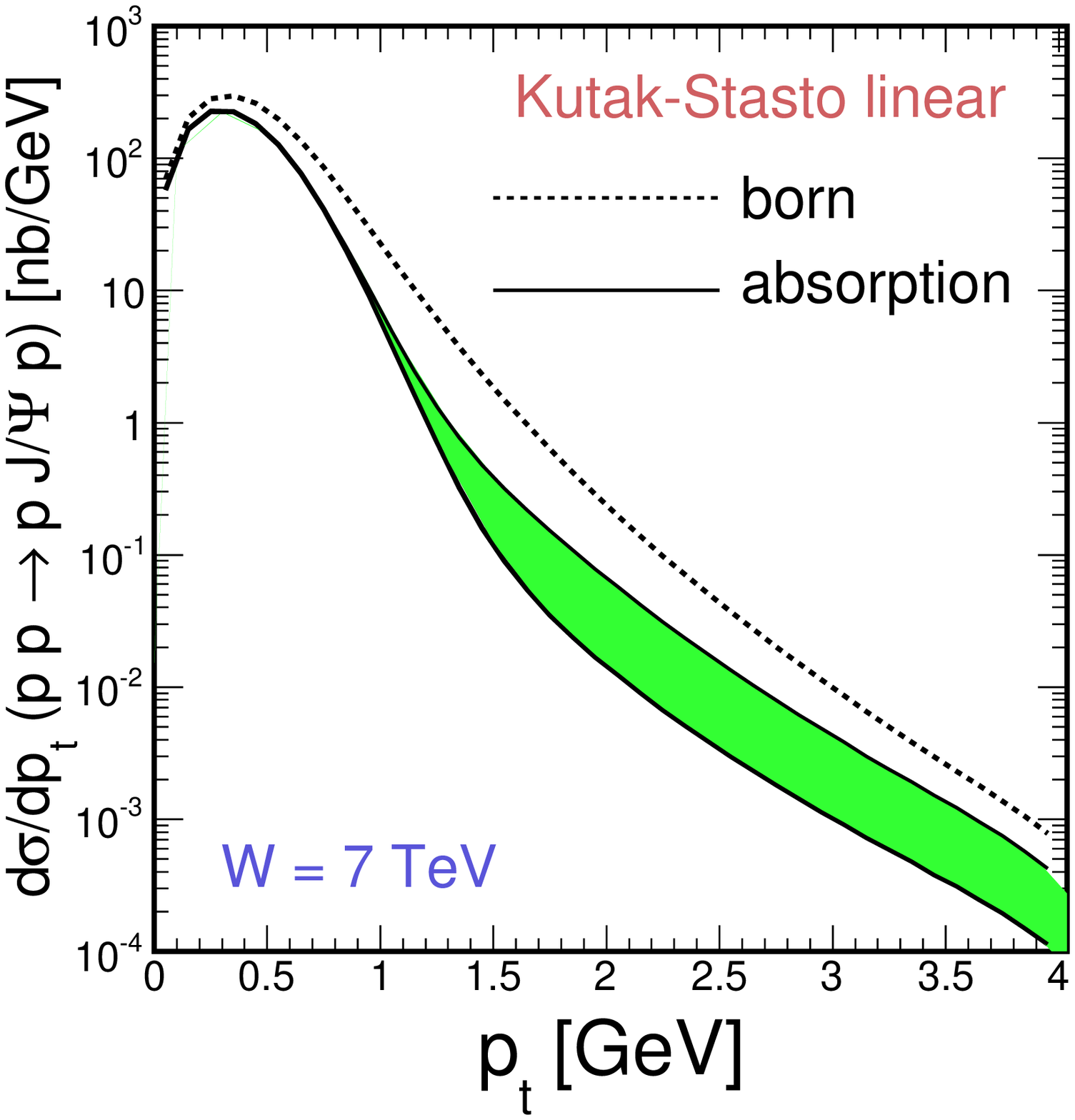}
\includegraphics[width=5.0cm]{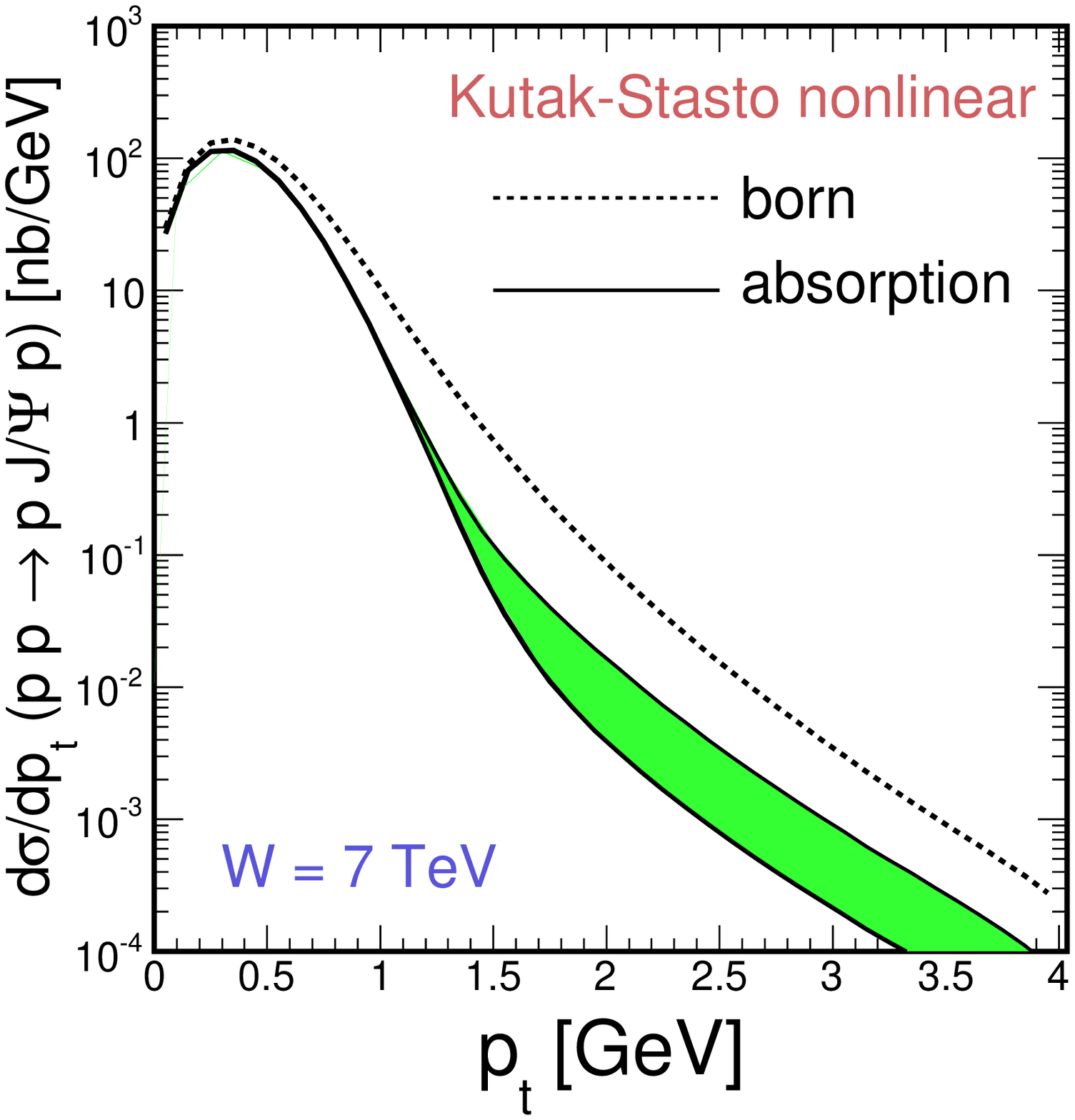}
\caption[*]{ 
$J/\psi$ transverse momentum distribution calculated with 
with absorption effects (solid line) and in the Born approximation
for $\sqrt{s}$ = 7 TeV.
The shaded (green online) band represents typical uncertainties 
in calculating absorption effects as described in the text.
\label{fig:dsig_dpt_absorption}
}
\end{center}
\end{figure}

How the absorption modifies the $J/\psi$ transverse momentum
distribution is shown in Fig.\ref{fig:dsig_dpt_absorption}.
The absorption leads to a strong damping at large $J/\psi$ transverse
momenta. This overcompensates the effect of inclusion of the tensor
electromagnetic coupling quantified by the Pauli electromagnetic
form factor.
In Fig.\ref{fig:dsig_dpt_absorption} we show both results with the standard
absorption (elastic rescattering only) as well as with the absorption
increased by a factor 1.4 to simulate inelastic (nucleon excitation) terms.

\begin{figure}[!htb] 
\begin{center}
\includegraphics[width=5.0cm]{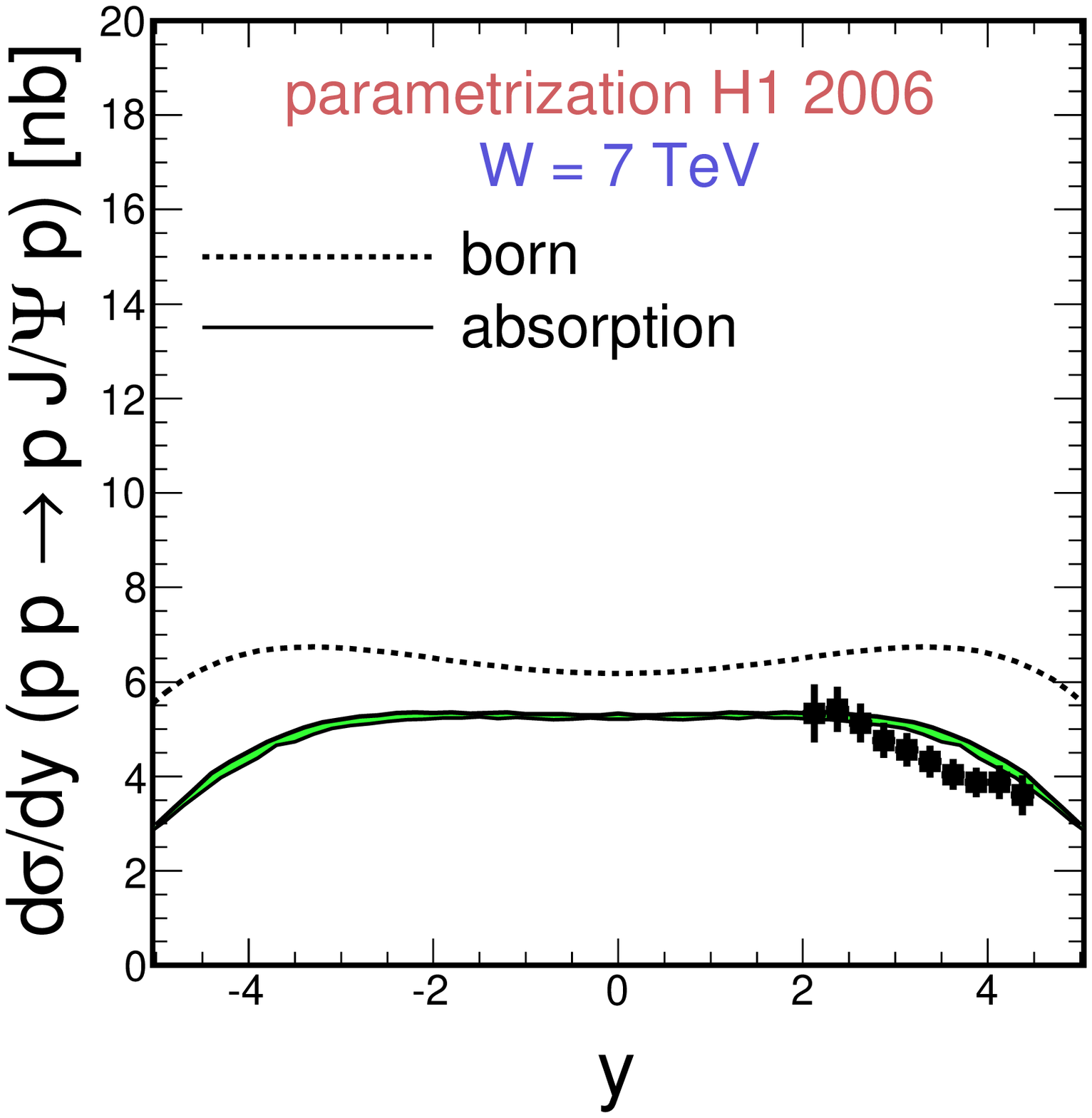}
\includegraphics[width=5.0cm]{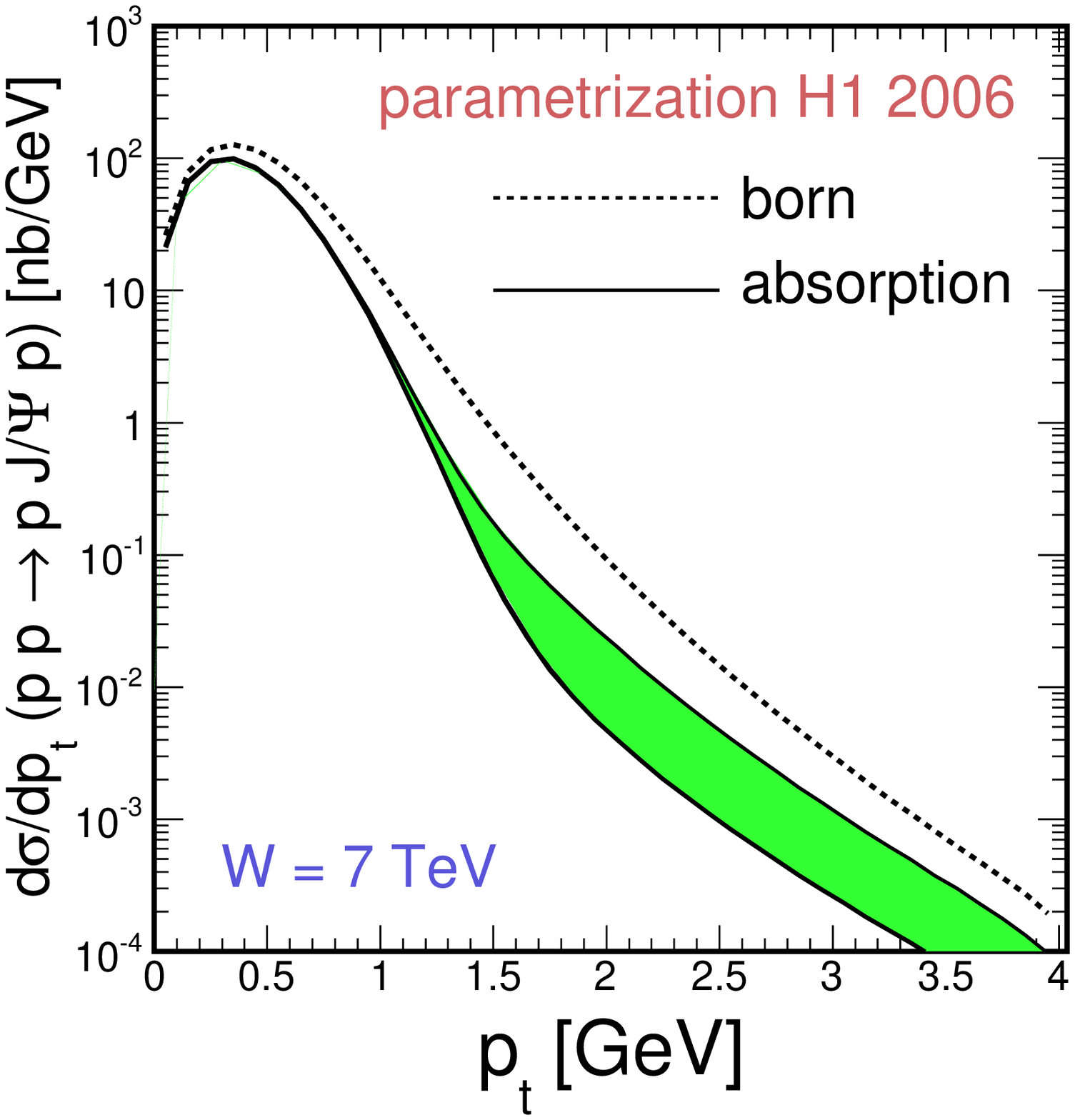}
\caption[*]{ 
$J/\psi$ rapidity and transverse momentum distributions calculated with 
the H1 parametrization \cite{H1_b,SS2007} of the elementary 
$\gamma p \to J/\psi p$ cross section for $\sqrt{s}$ = 7 TeV.
The new LHCb data points \cite{LHCb_second} are shown for comparison.
\label{fig:dsig_dy_parametrization}
}
\end{center}
\end{figure}

For completeness in Fig.\ref{fig:dsig_dy_parametrization} we present
rapidity (left panel) and transverse momentum (right panel) distributions 
obtained using H1 parametrization \cite{H1_b} of the elementary
$\gamma p \to J/\psi p$ cross section (see Fig.\ref{fig:sig_tot_W}).
A good agreement with the LHCb data precludes drawing definite
conclusion about onset of saturation.

\subsection{$\psi'$ production}

\begin{figure}[!htb] 
\begin{center}
\includegraphics[height=6.0cm]{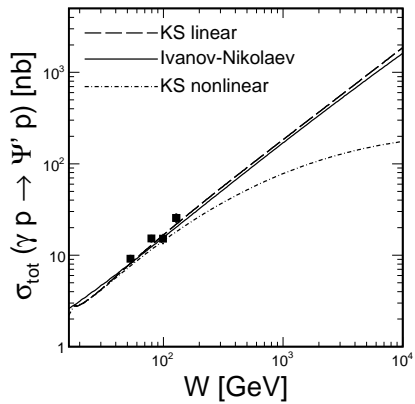}
\caption[*]{ 
Total cross section for the $\gamma p \to \psi' p$ as a function of
the subsystem energy together with the HERA data and pseudodata
obtained by the LHCb collaboration. Three disfferent UGDFs have been
used: Ivanov-Nikolaev (solid), Kutak-Stasto linear (dashed) and 
Kutak-Stasto nonlinear (dash-dotted). The experimental data are from
Ref. \cite{HERA_new}.
\label{fig:sig_tot_W_2S}
}
\end{center}
\end{figure}

Now we shall proceed to the production of the excited charmonium state 
$\psi'$. In Fig. \ref{fig:sig_tot_W_2S} we present total cross section
for the $\gamma p \to \psi' p$ as a function of collision energy 
for the different UGDFs considered here. Almost all the UGDFs
provide good description of new HERA data \cite{HERA_new}. 
The description seems better than in recent analysis in 
Ref.\cite{JMRT2013_Psi_2S}
where the collinear gluon distribution fitted to the production
of $J/\psi$ \cite{JMRT2013} was used.
Part of the success is due to explicit use of light cone wave
functions which are not explicit in the collinear approximation
as discussed above.

\begin{figure}[!htb] 
\begin{center}
\includegraphics[width=5.0cm]{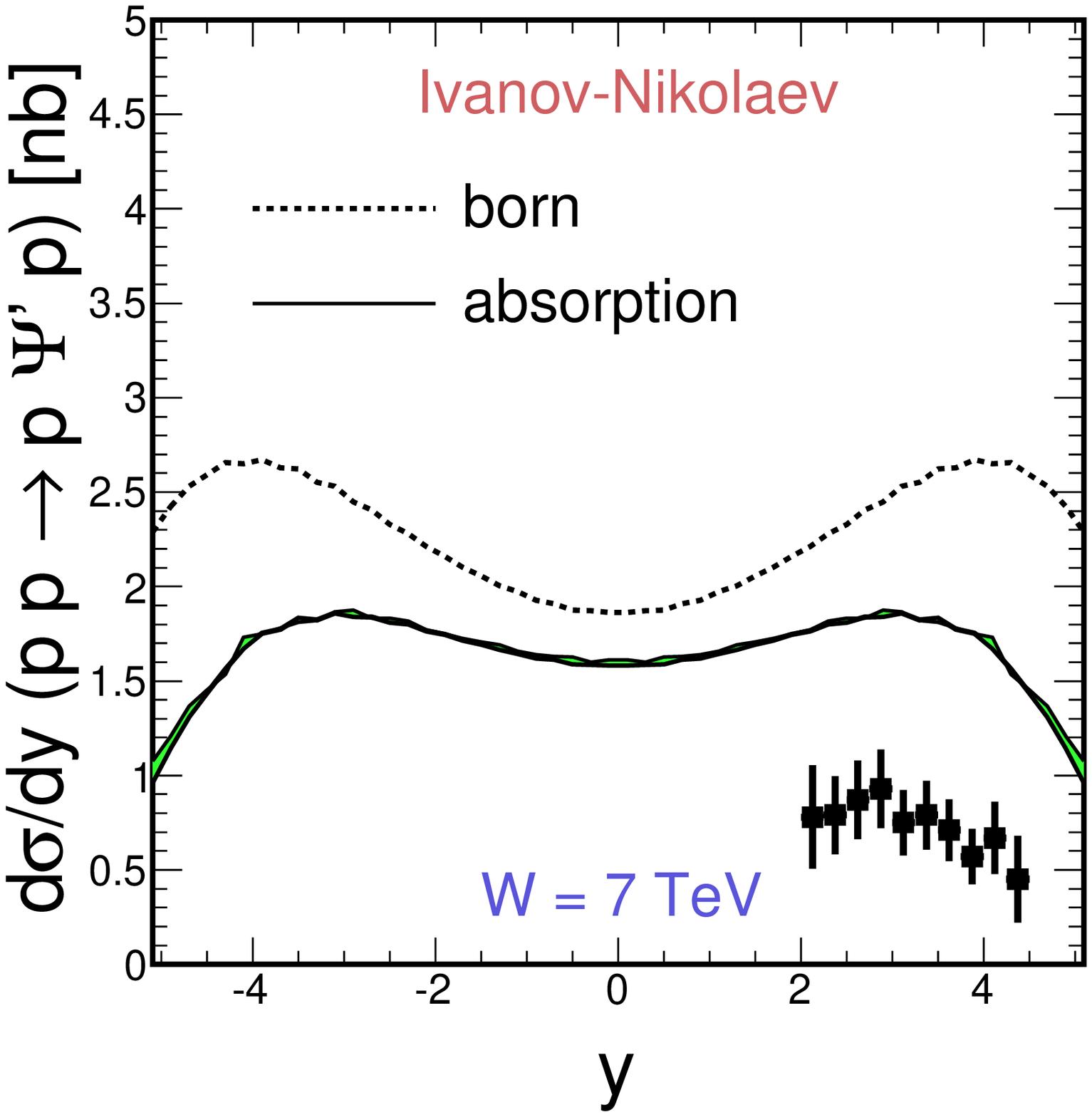}
\includegraphics[width=5.0cm]{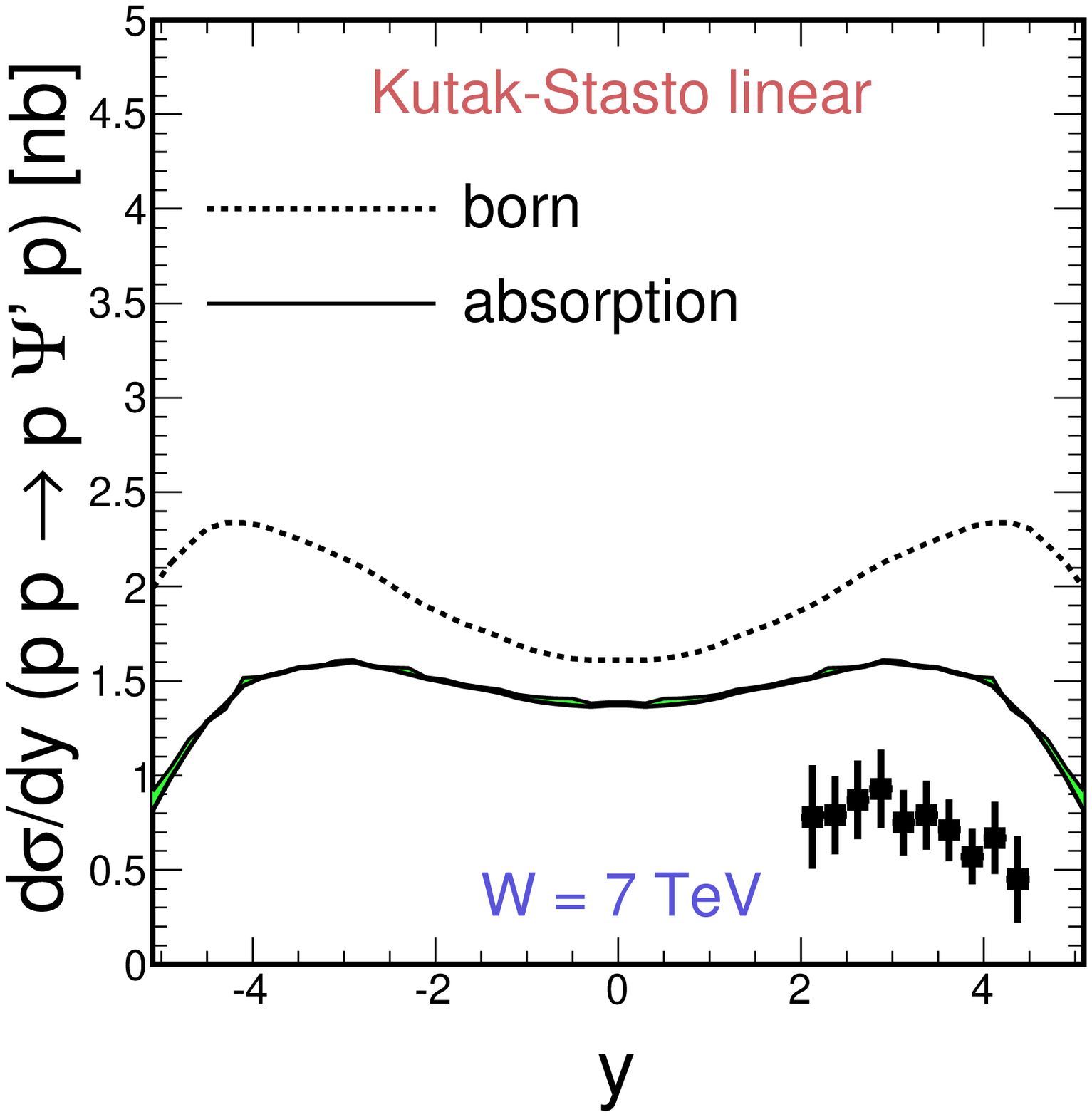}
\includegraphics[width=5.0cm]{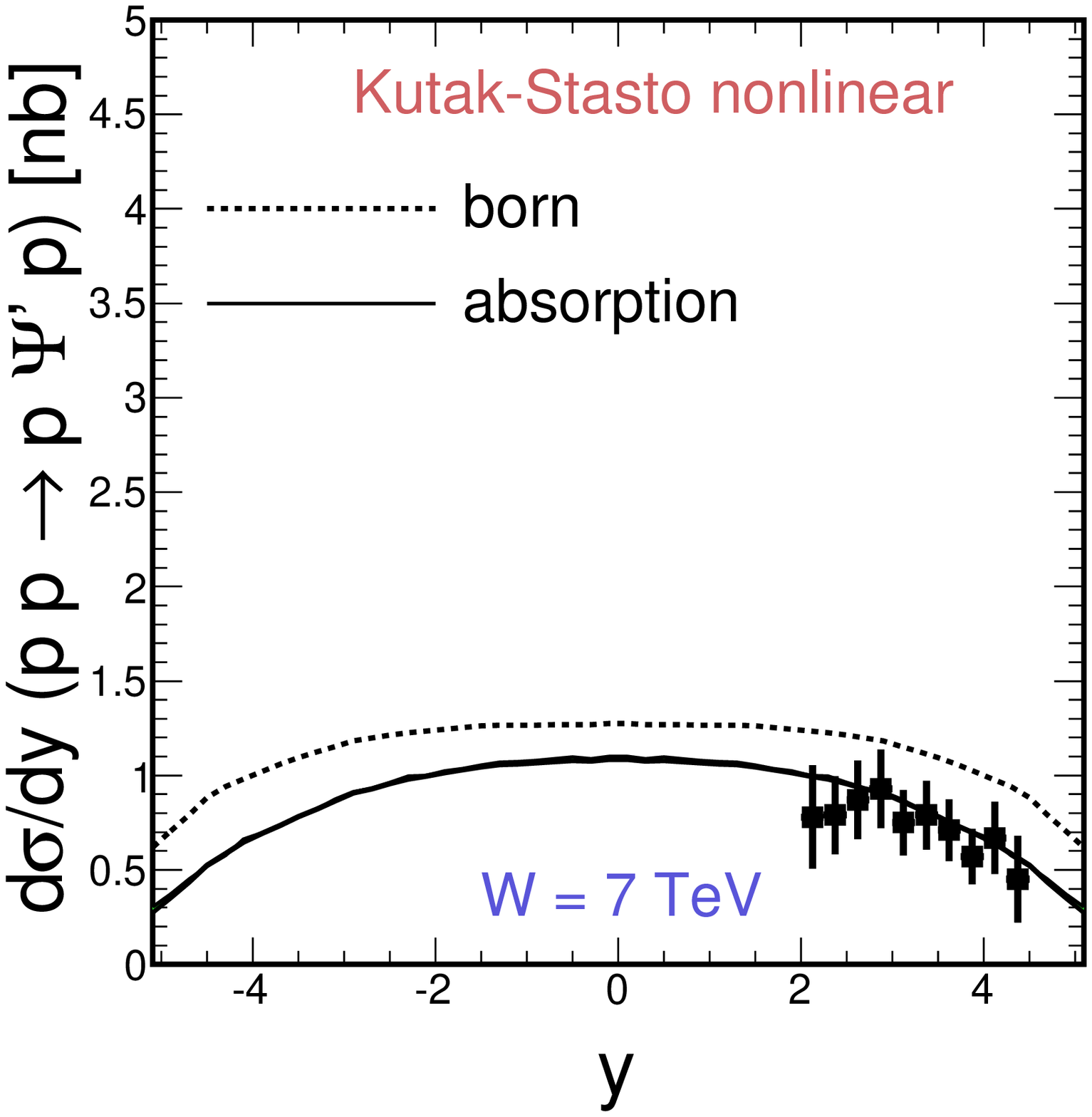}

\caption[*]{ 
Rapidity distribution of $\psi'$ calculated with inclusion of 
absorption effects (solid line), compared with the result when 
absorption effects are ignored (dotted line) for $\sqrt{s}$ = 7 TeV.
The new LHCb data points \cite{LHCb_second} are shown for comparison.
\label{fig:dsig_dy_2S_absorption}
}
\end{center}
\end{figure}

Now we turn to hadronic collisions. In
Fig.\ref{fig:dsig_dy_2S_absorption} we present our predictions for
rapidity distributions of $\psi'$ again for different UGDFs.
Results of our calculations are compared with recent LHCb data 
\cite{LHCb_second}.

\begin{figure}[!htb] 
\begin{center}
\includegraphics[width=5.0cm]{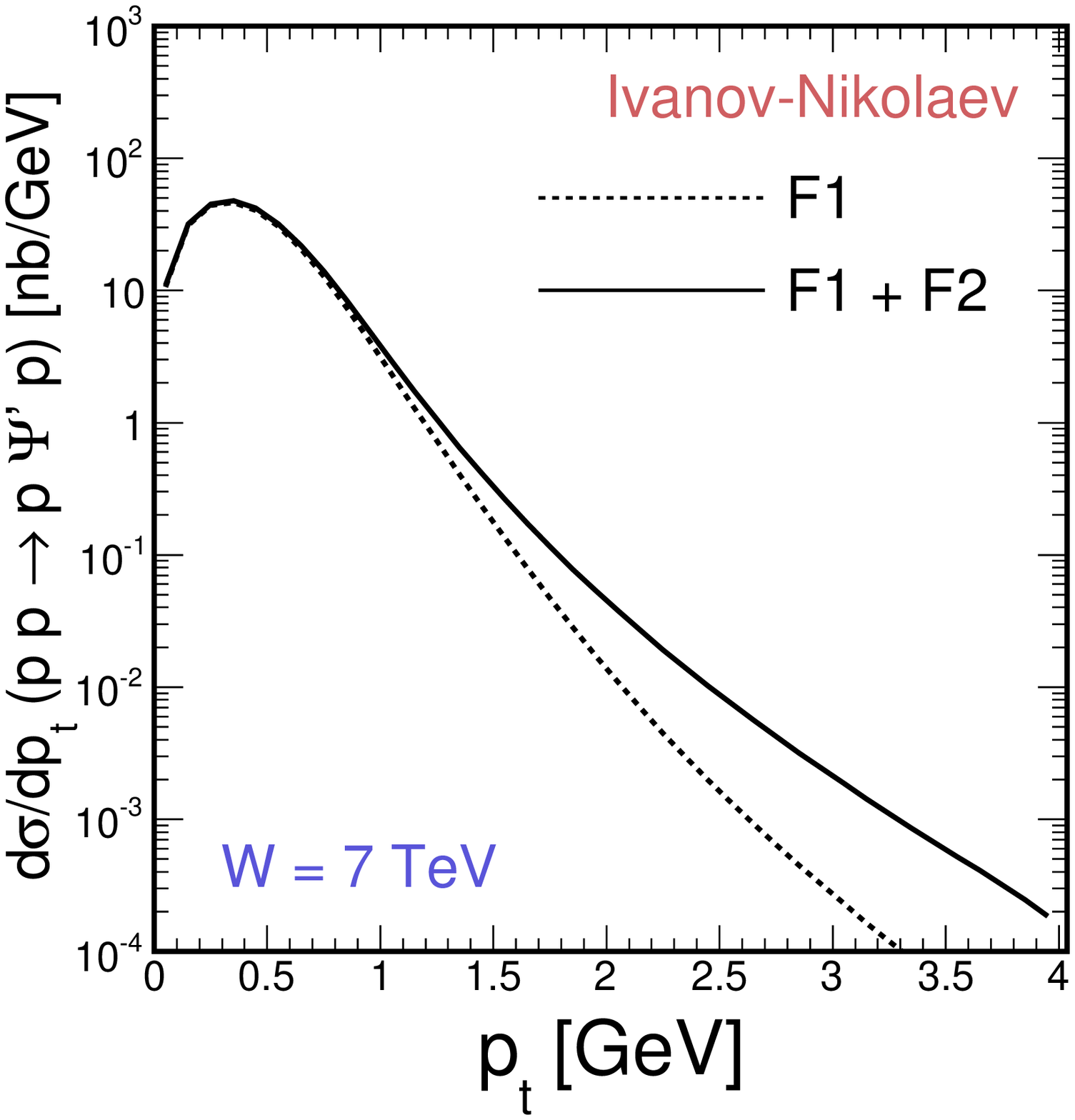}
\includegraphics[width=5.0cm]{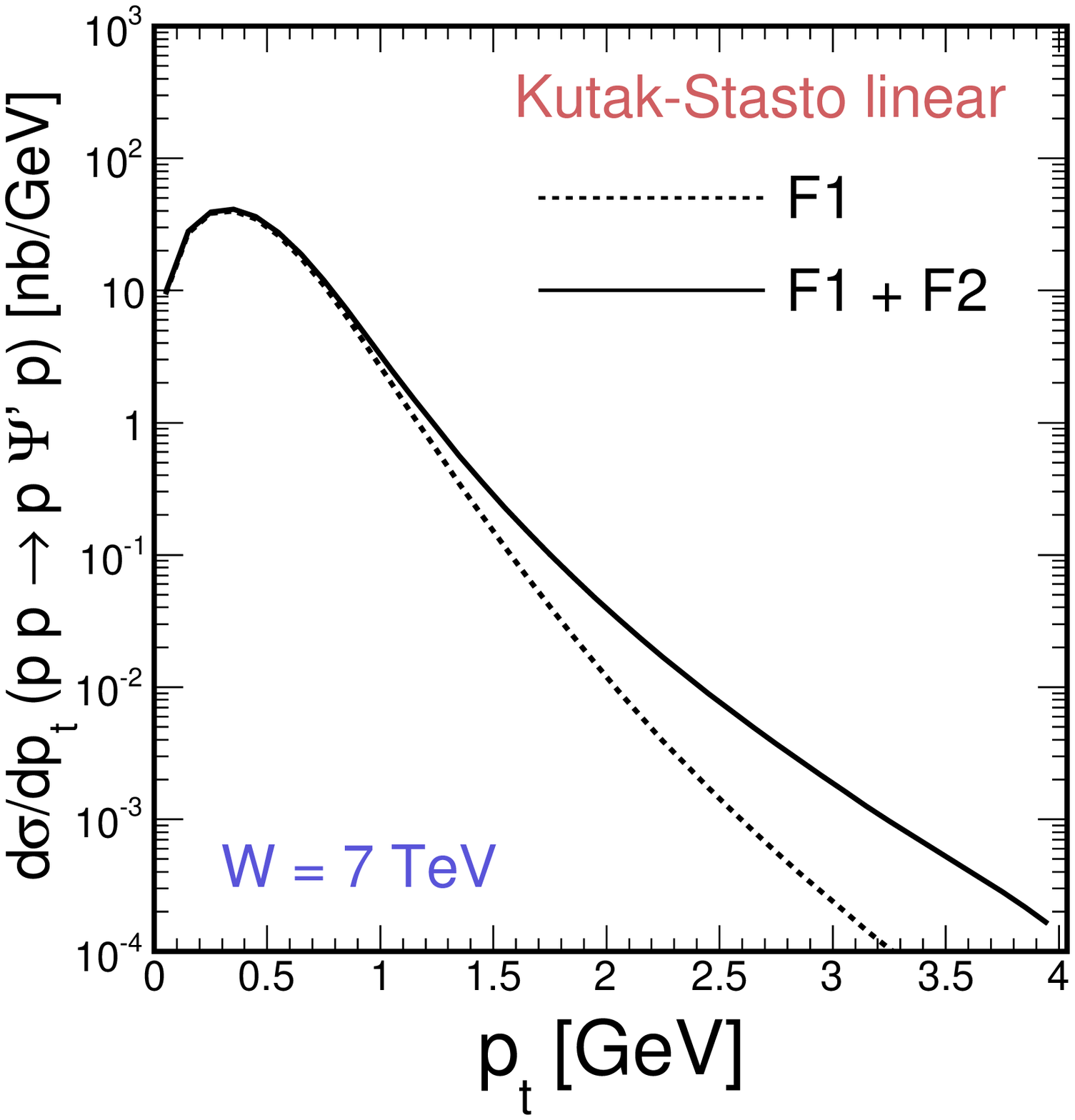}
\includegraphics[width=5.0cm]{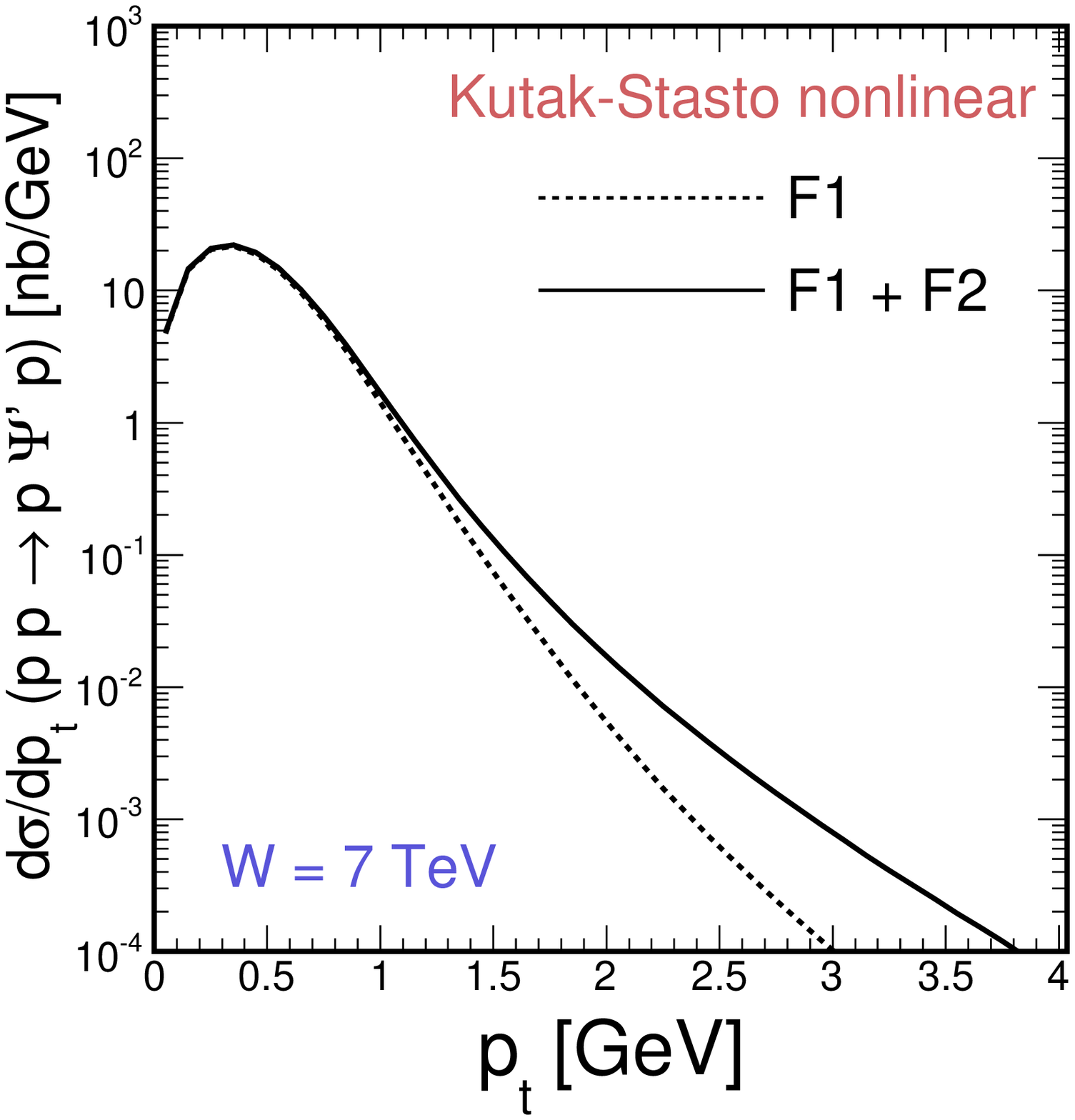}
\caption[*]{ 
$\psi'$ transverse momentum distribution calculated 
in the Born approximation with and without including Pauli
electromagnetic form factor for $\sqrt{s}$ = 7 TeV.
\label{fig:dsig_dpt_2S_Born}
}
\end{center}
\end{figure}

The role of Pauli electromagnetic form factor is quantified
in Fig.\ref{fig:dsig_dpt_2S_Born}. As for the ground state $J/\psi$
the tensor coupling enhances the cross section at large meson transverse
momenta.

\begin{figure}[!htb] 
\begin{center}
\includegraphics[width=5.0cm]{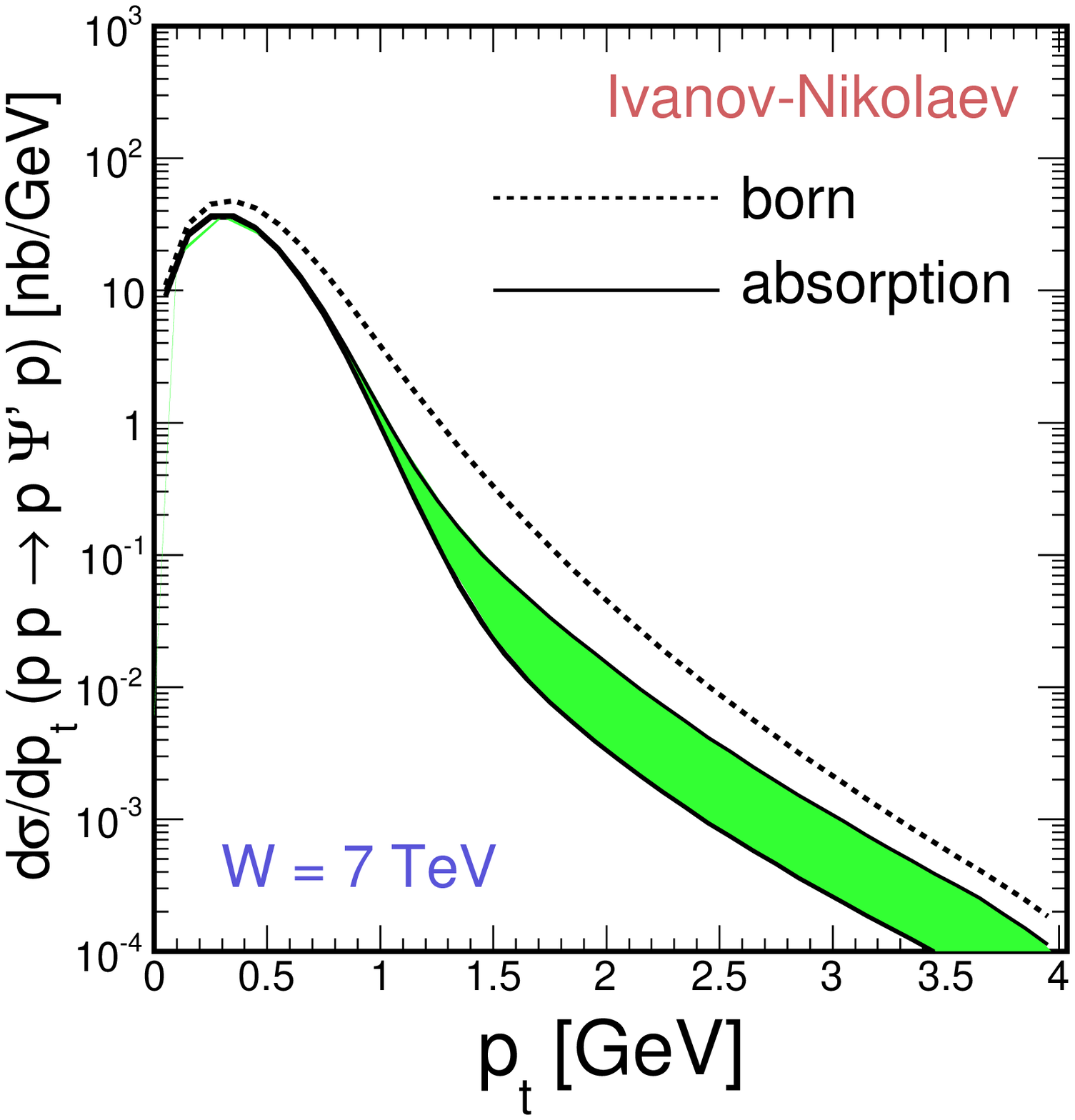}
\includegraphics[width=5.0cm]{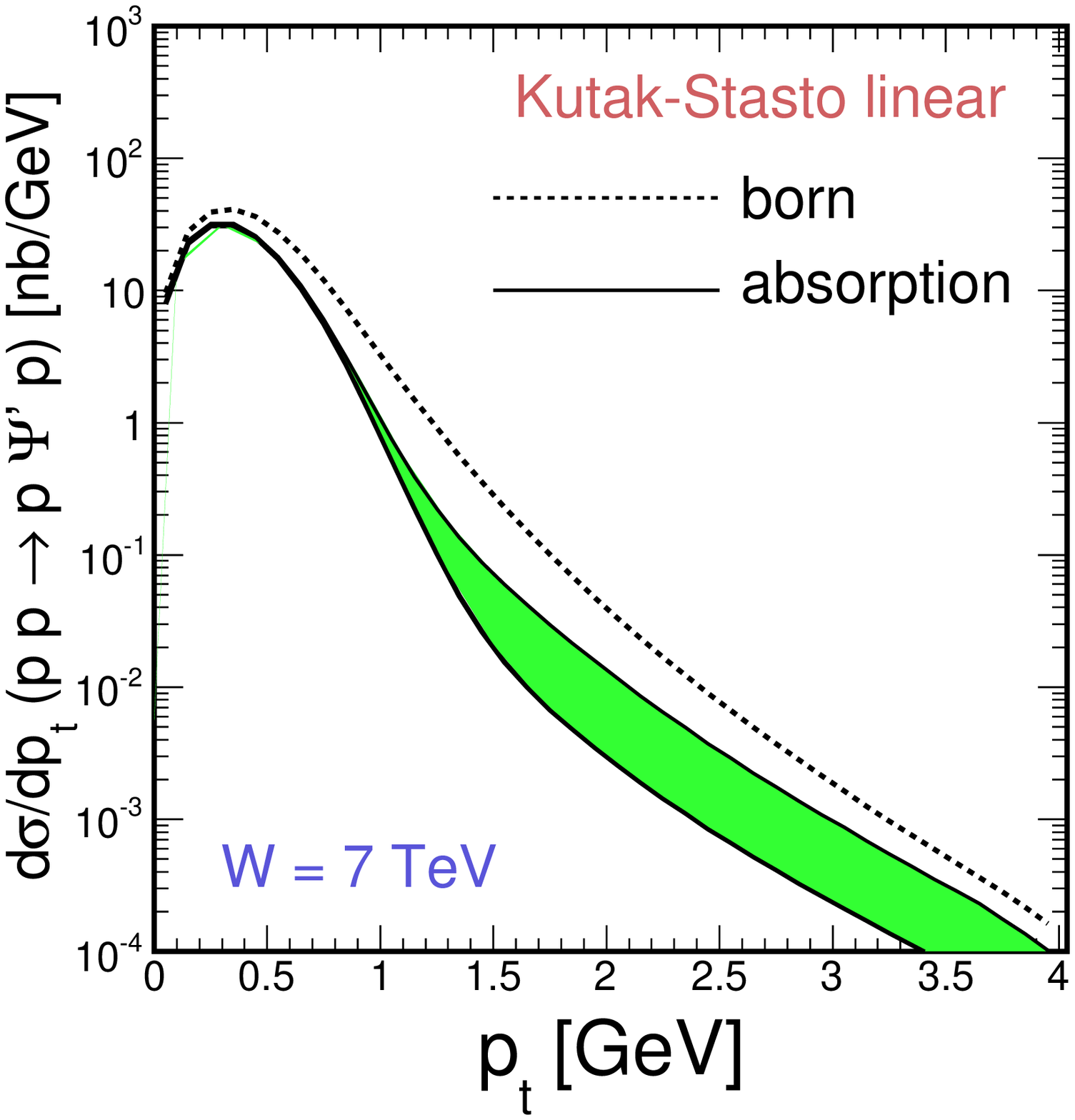}
\includegraphics[width=5.0cm]{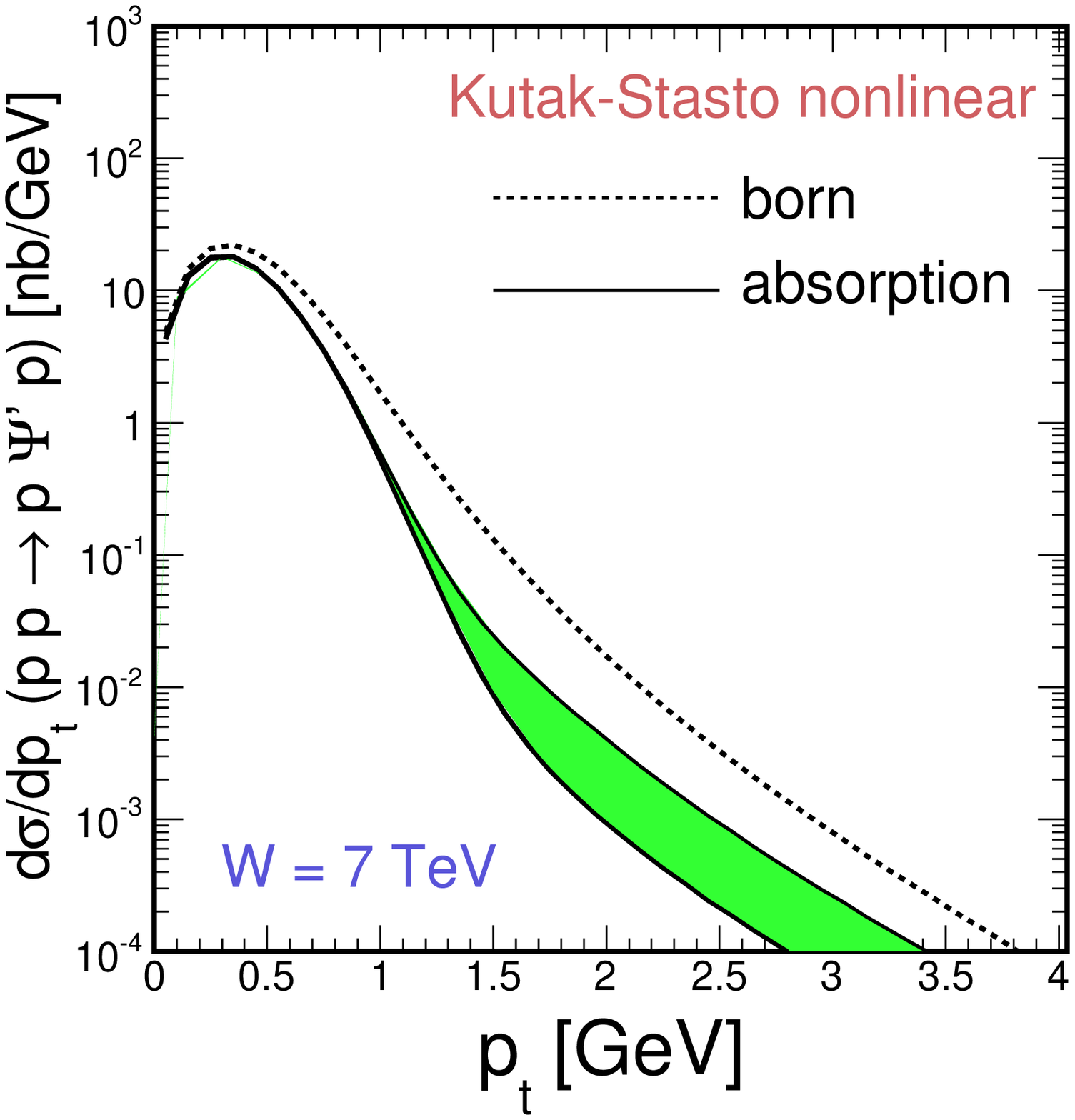}
\caption[*]{ 
$\psi'$ transverse momentum distribution calculated with 
absorption effects (solid line) and in the Born approximation
(dashed line) for $\sqrt{s}$ = 7 TeV.
\label{fig:dsig_dpt_2S_absorption}
}
\end{center}
\end{figure}

The role of absorption effects is discussed in 
Fig.\ref{fig:dsig_dpt_2S_absorption}. The Born results are
shown for comparison. The absorption effects lead to strong damping
of of the cross section at large $p_t$'s. This is a region where 
odderon exchange may show up.

\subsection{$J/\psi$ and $\psi'$ production at the Tevatron}

\begin{figure}[!htb] 
\begin{center}
\includegraphics[width=5.0cm]{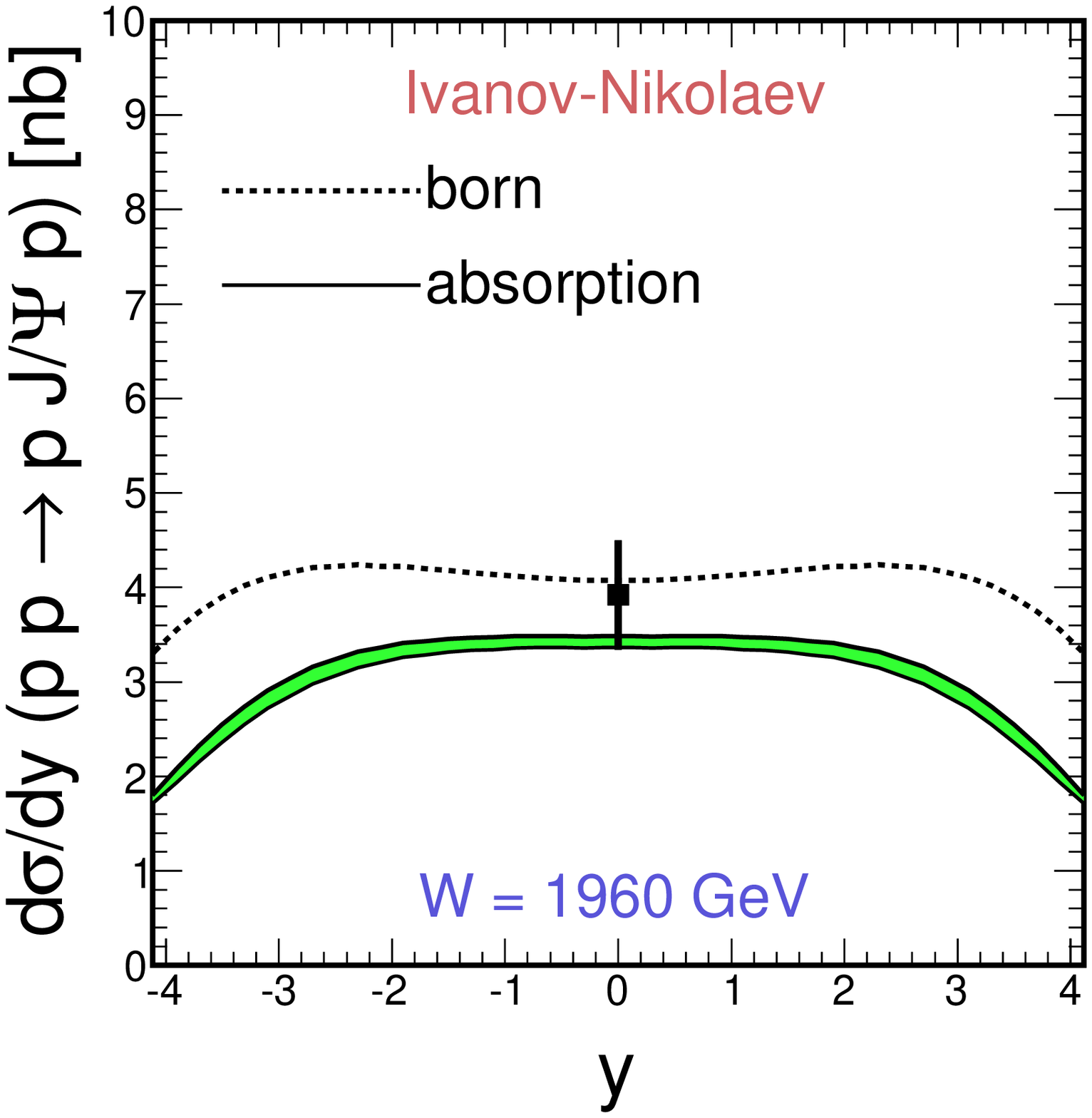}
\includegraphics[width=5.0cm]{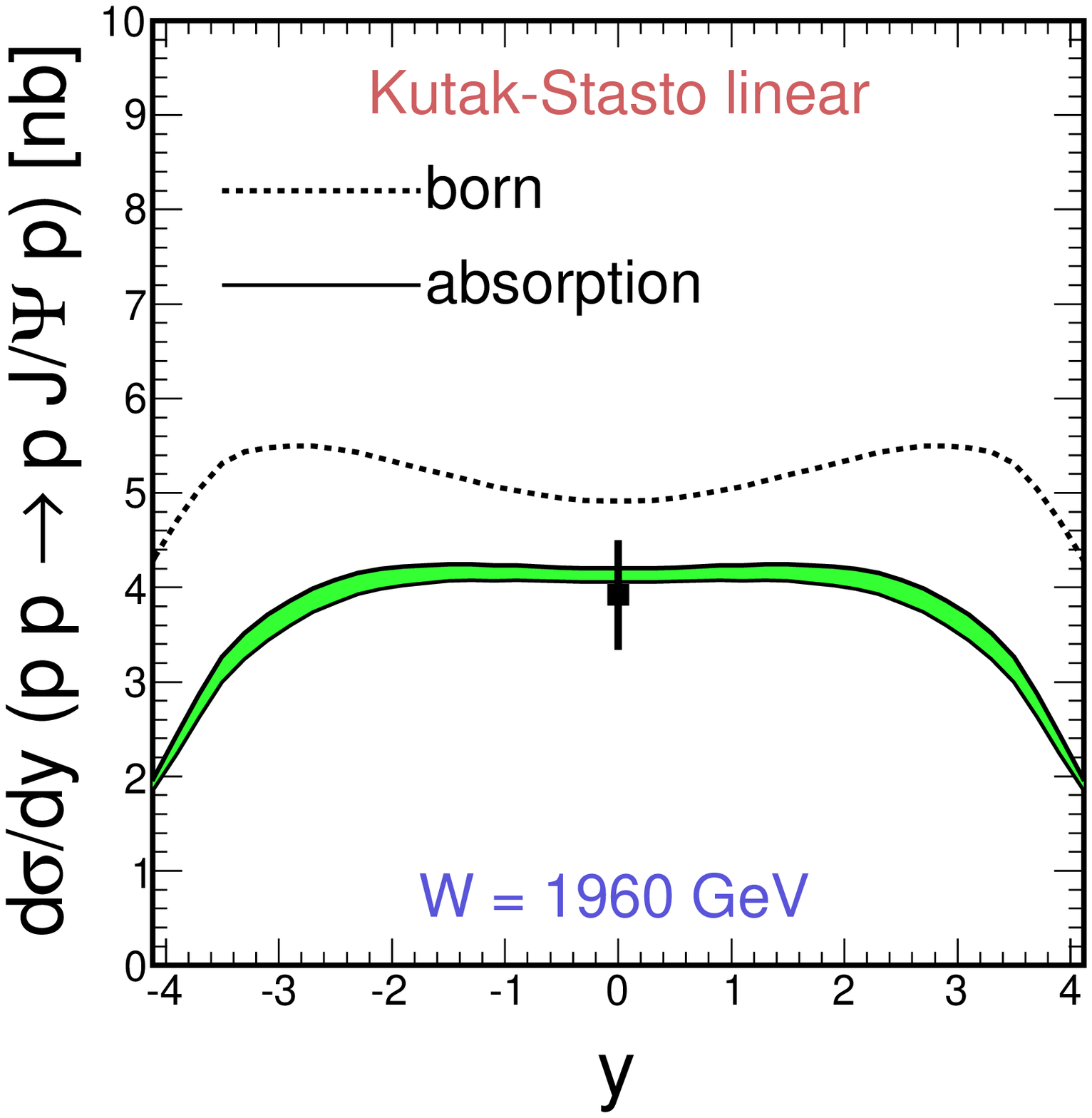}
\includegraphics[width=5.0cm]{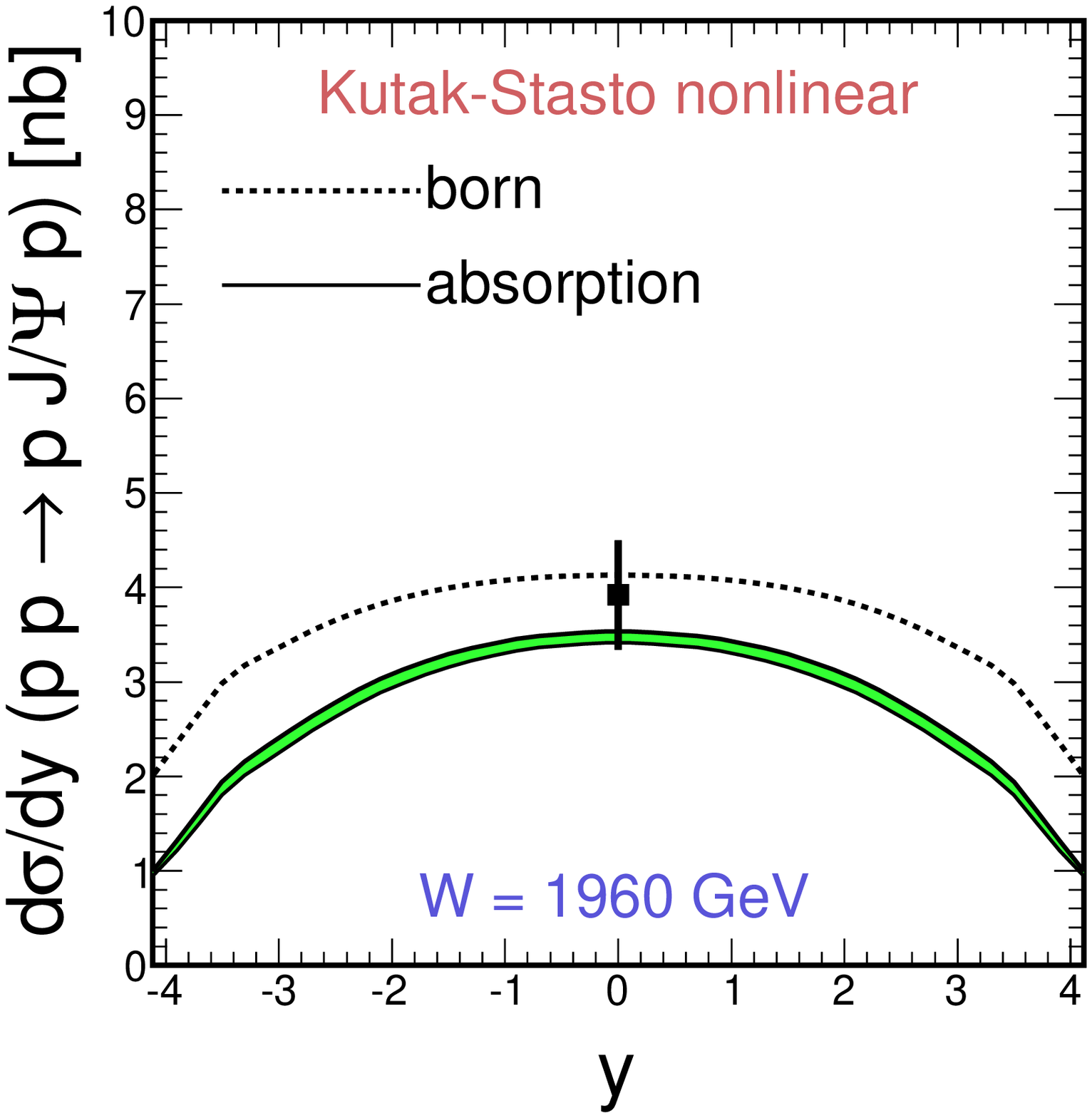}

\caption[*]{ 
$J/\psi$ rapidity distribution calculated with inclusion of 
absorption effects (solid line), compared with the Born result 
(dashed line) for $\sqrt{s}$ = 1.96 TeV. 
The CDF data point \cite{CDF} is shown for comparison.
\label{fig:dsig_dy_absorption_Tev}
}
\end{center}
\end{figure}

In this section for completeness we present also results for the Tevatron. 
We repeat the same presentation as for the LHC. 
In Fig.\ref{fig:dsig_dy_absorption_Tev} we show distribution 
in rapidity of $J/\psi$. All UGDFs considered here describe 
the CDF data point if the absorption effects are included.

\begin{figure}[!htb] 
\begin{center}
\includegraphics[width=5.0cm]{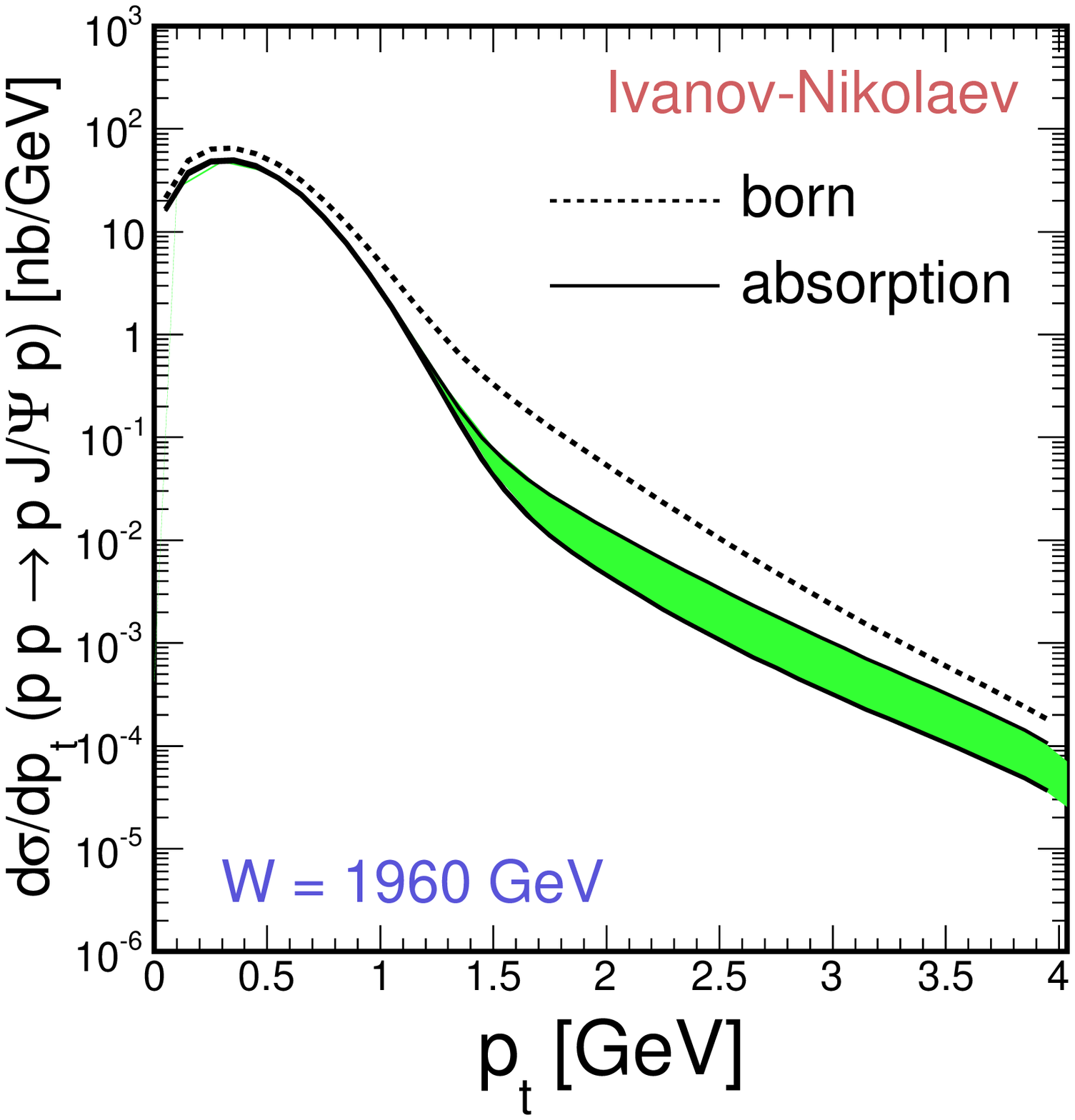}
\includegraphics[width=5.0cm]{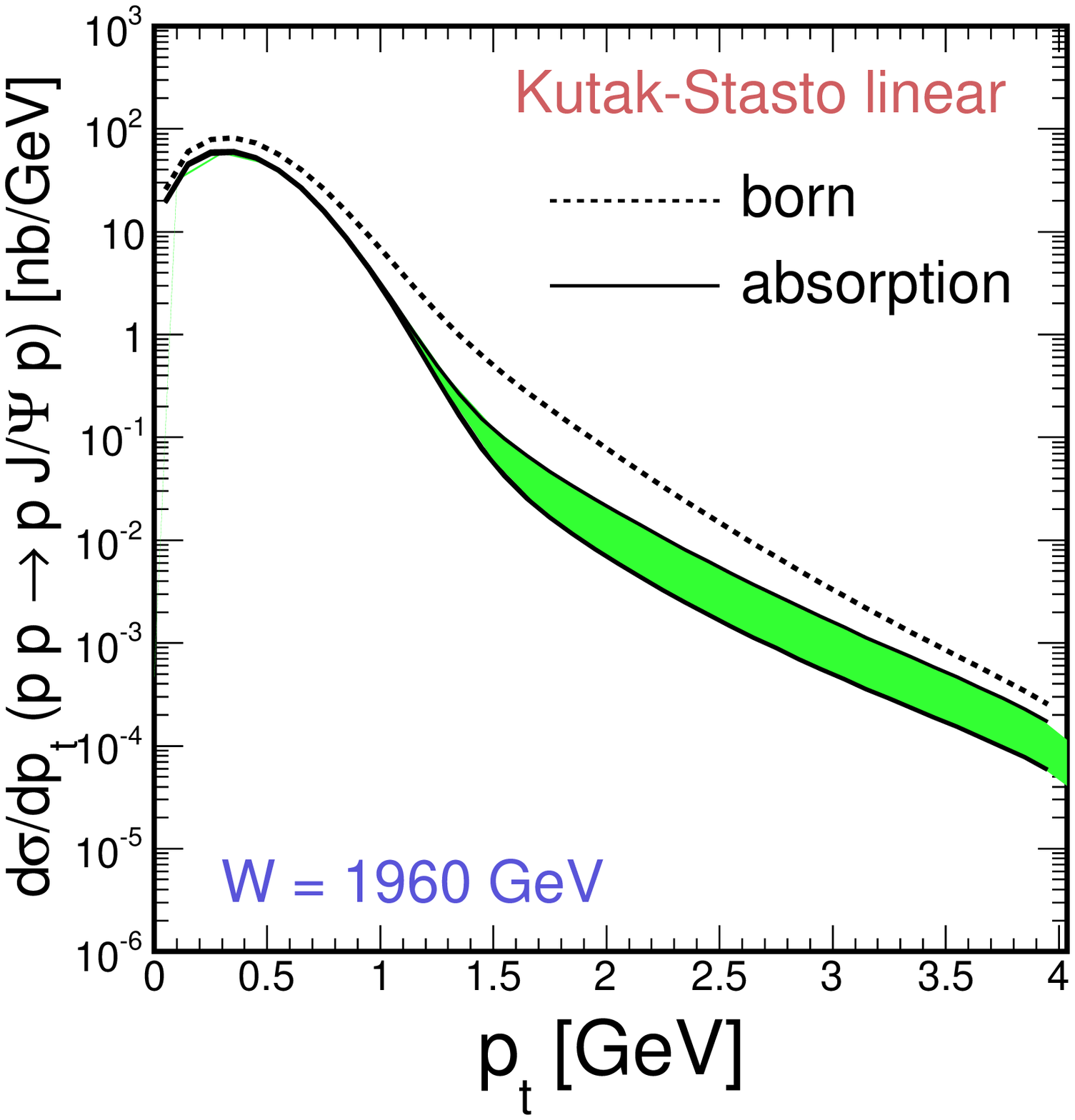}
\includegraphics[width=5.0cm]{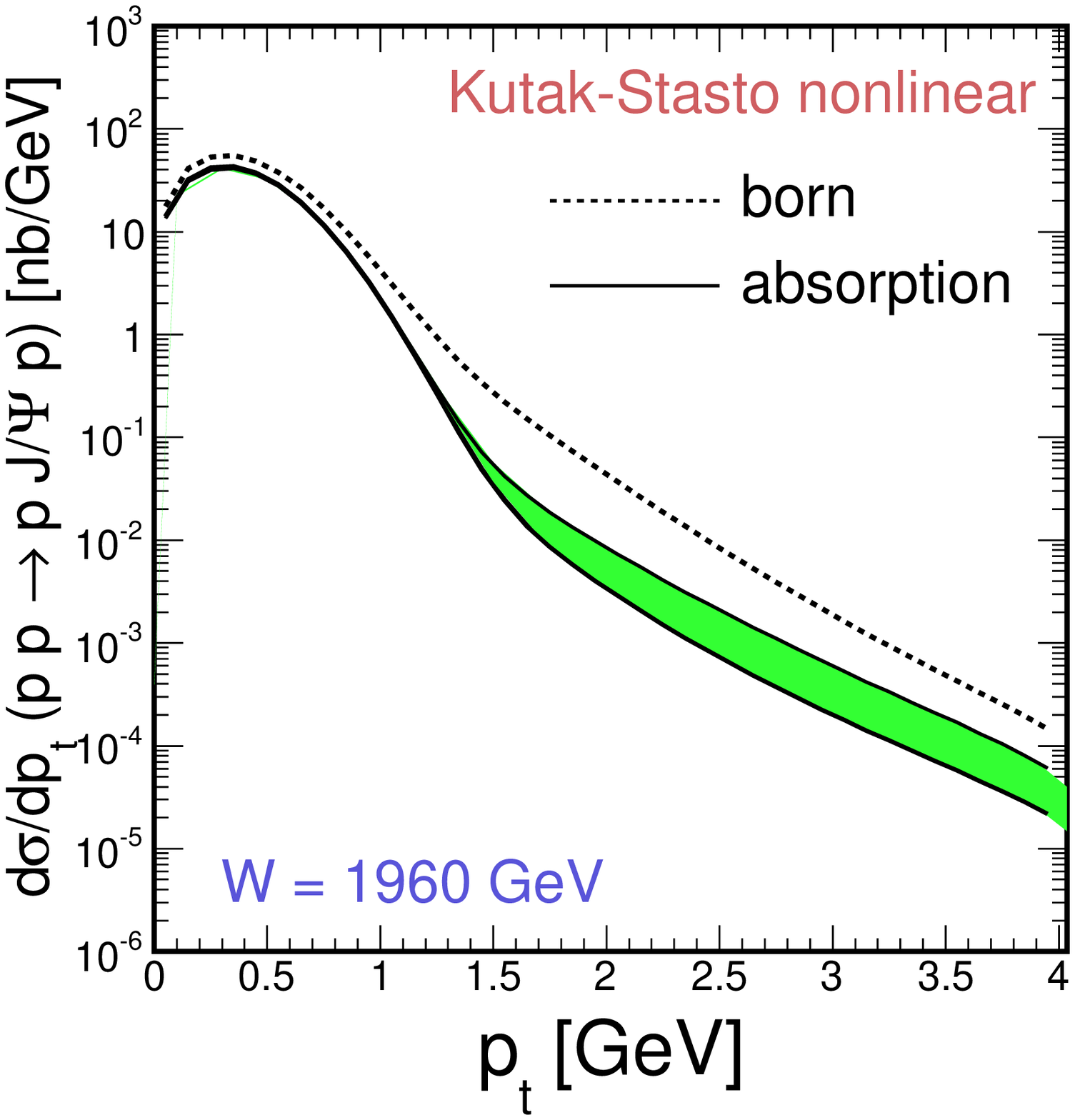}
\caption[*]{ 
$J/\psi$ transverse momentum distribution calculated with absorption effects
(solid line) and in the Born approximation (dashed line).
\label{fig:dsig_dpt_absorption_Tev}
}
\end{center}
\end{figure}

In the next figure (Fig.\ref{fig:dsig_dpt_Born_Tev}) we demonstrate 
the role of the Pauli form factor on the $J/\psi$ transverse momentum
distribution. The distributions for the Tevatron (larger $x$ values
of the gluon distribution) 
drop much quicker than those for the LHC 
(lower $x$ values of the gluon distribution).
The role of absorption is presented in 
Fig.\ref{fig:dsig_dpt_absorption_Tev}.

\begin{figure}[!htb] 
\begin{center}
\includegraphics[width=5.0cm]{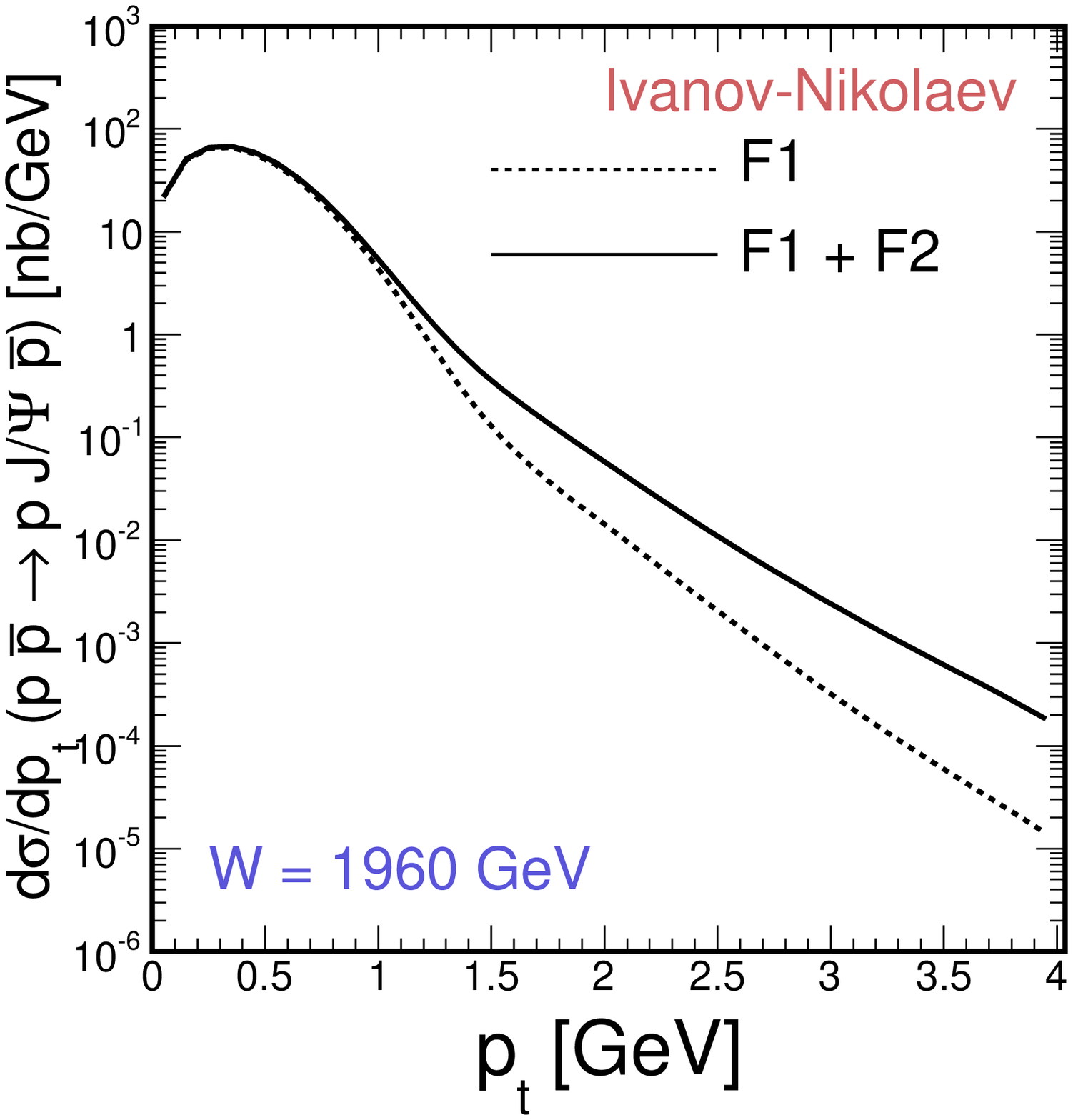}
\includegraphics[width=5.0cm]{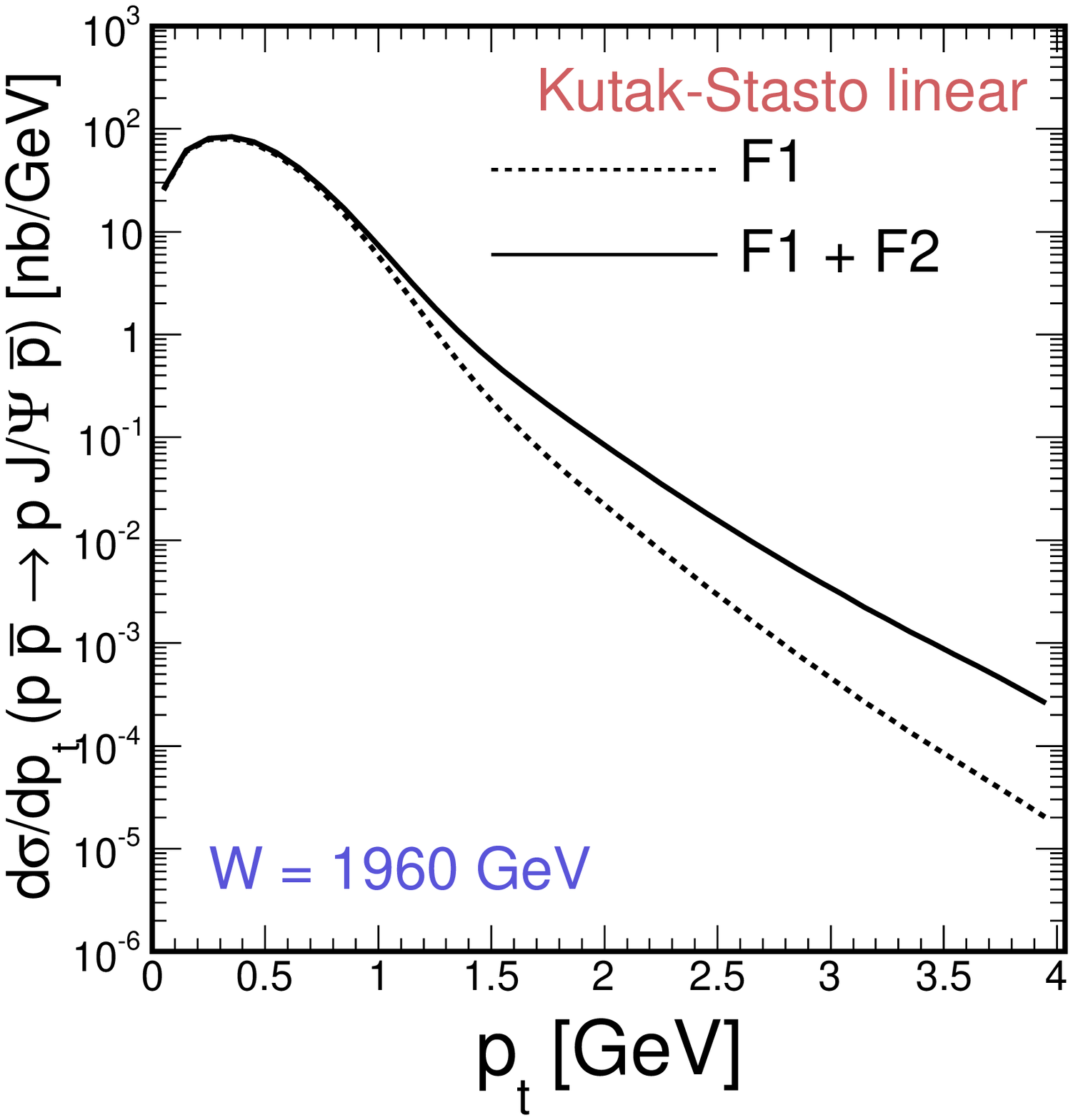}
\includegraphics[width=5.0cm]{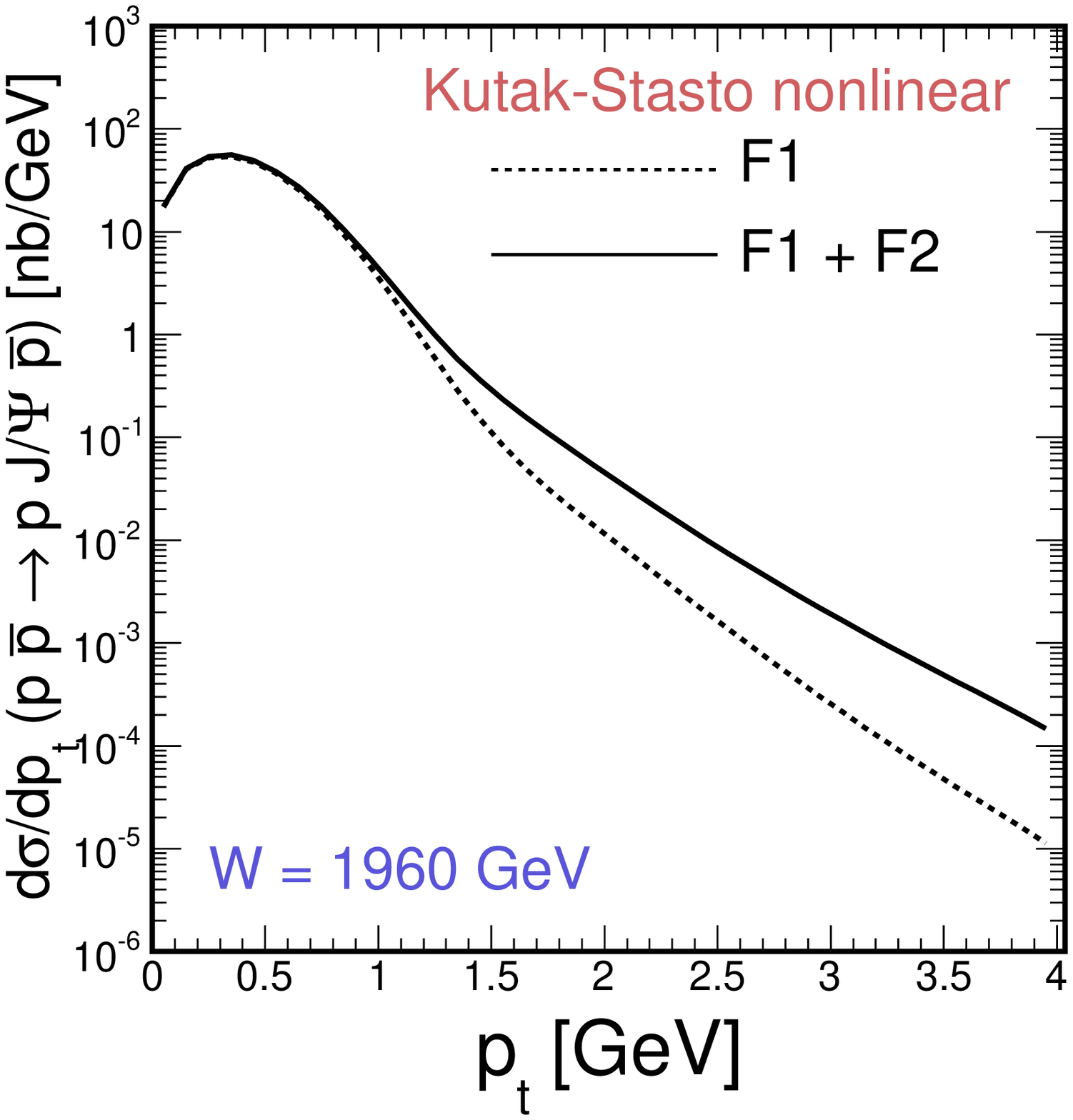}
\caption[*]{ 
$J/\psi$ transverse momentum distribution calculated with 
the Born amplitudes for three different UGDFs from the literature
for $\sqrt{s}$ = 1.96 TeV.
The dashed lines include contributions with Dirac $F_1$ electromagnetic 
form factor and the solid lines include in addition Pauli $F_2$ 
electromagnetic form factor.
\label{fig:dsig_dpt_Born_Tev}
}
\end{center}
\end{figure}

\begin{figure}[!htb] 
\begin{center}
\includegraphics[width=5.0cm]{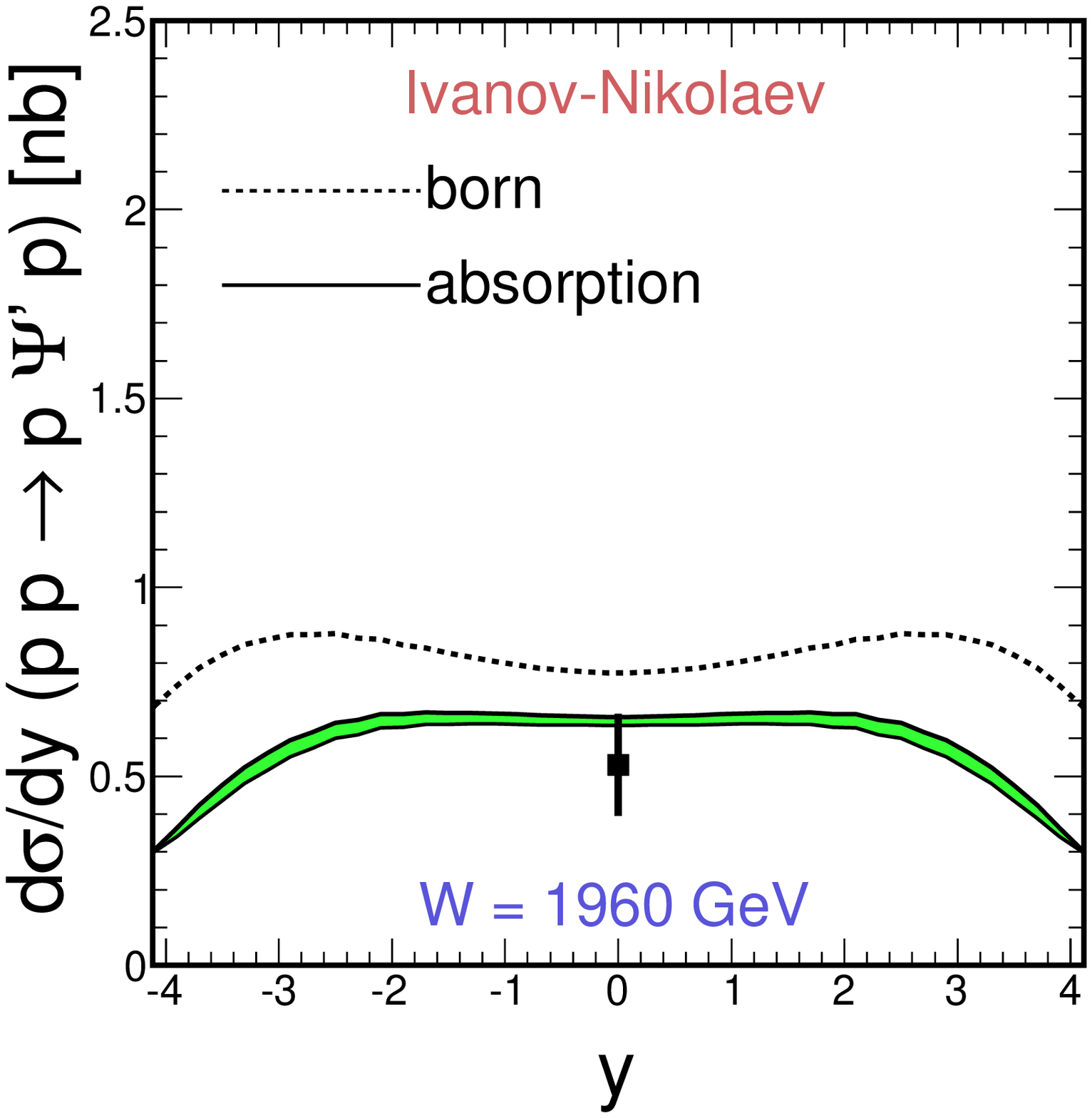}
\includegraphics[width=5.0cm]{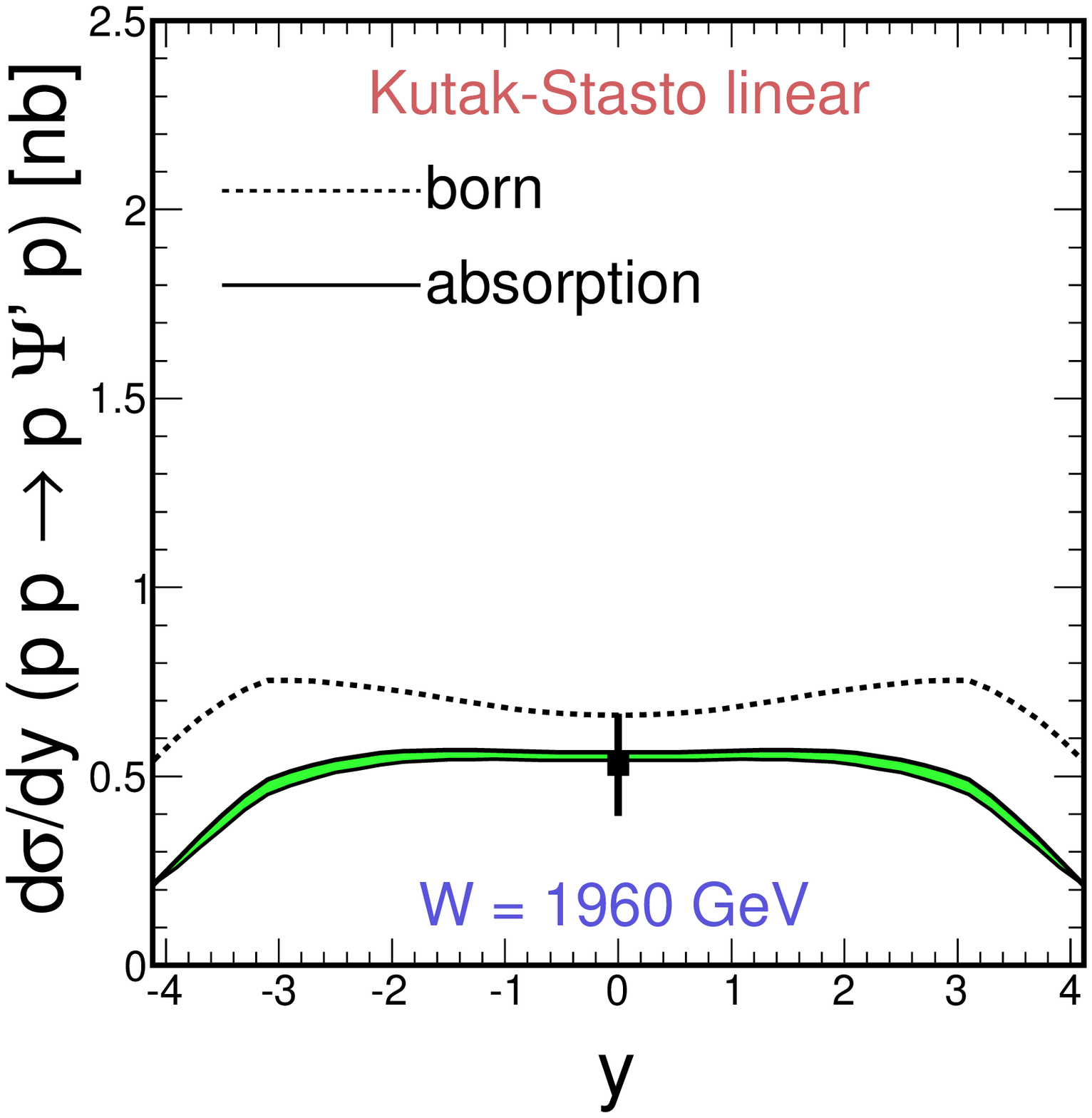}
\includegraphics[width=5.0cm]{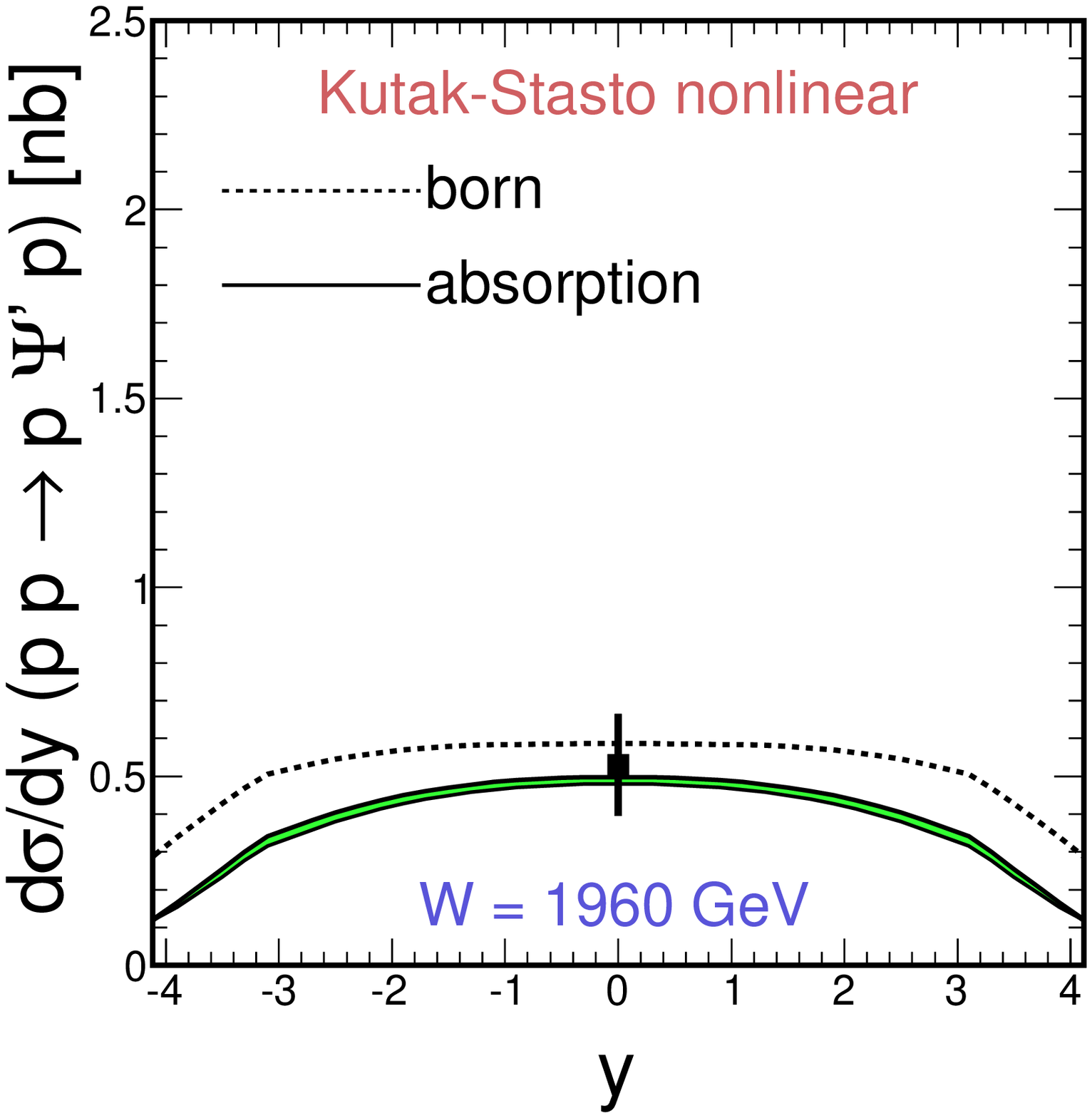}

\caption[*]{ 
Rapidity distribution of $\psi'$ calculated with inclusion of 
absorption effects (solid line), compared with the result when 
absorption effects are ignored (dotted line) for $\sqrt{s}$ = 1.96 TeV.
The CDF data point \cite{CDF} is shown for comparison.
\label{fig:dsig_dy_2S_absorption_Tev}
}
\end{center}
\end{figure}

\begin{figure}[!htb] 
\begin{center}
\includegraphics[width=5.0cm]{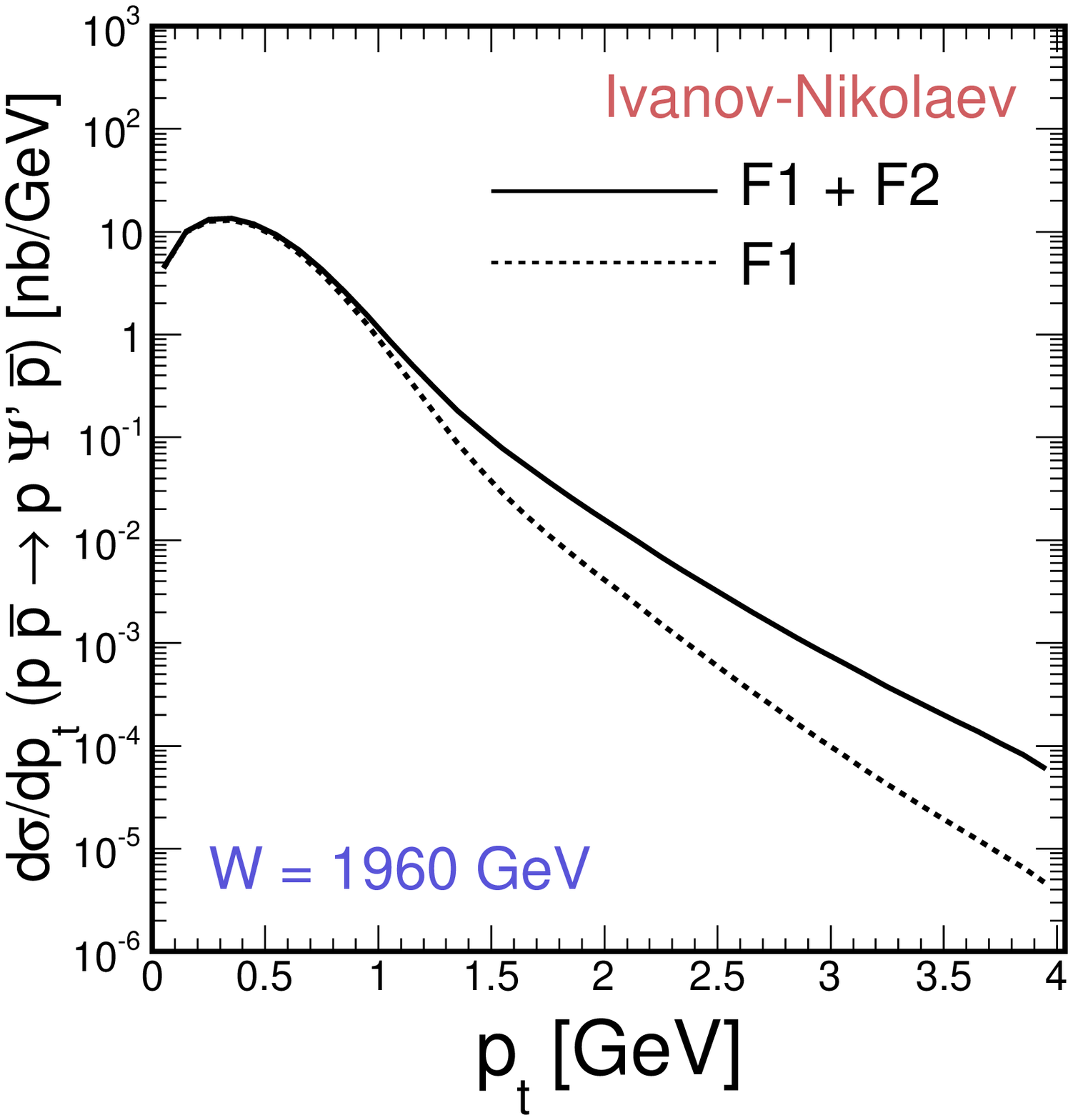}
\includegraphics[width=5.0cm]{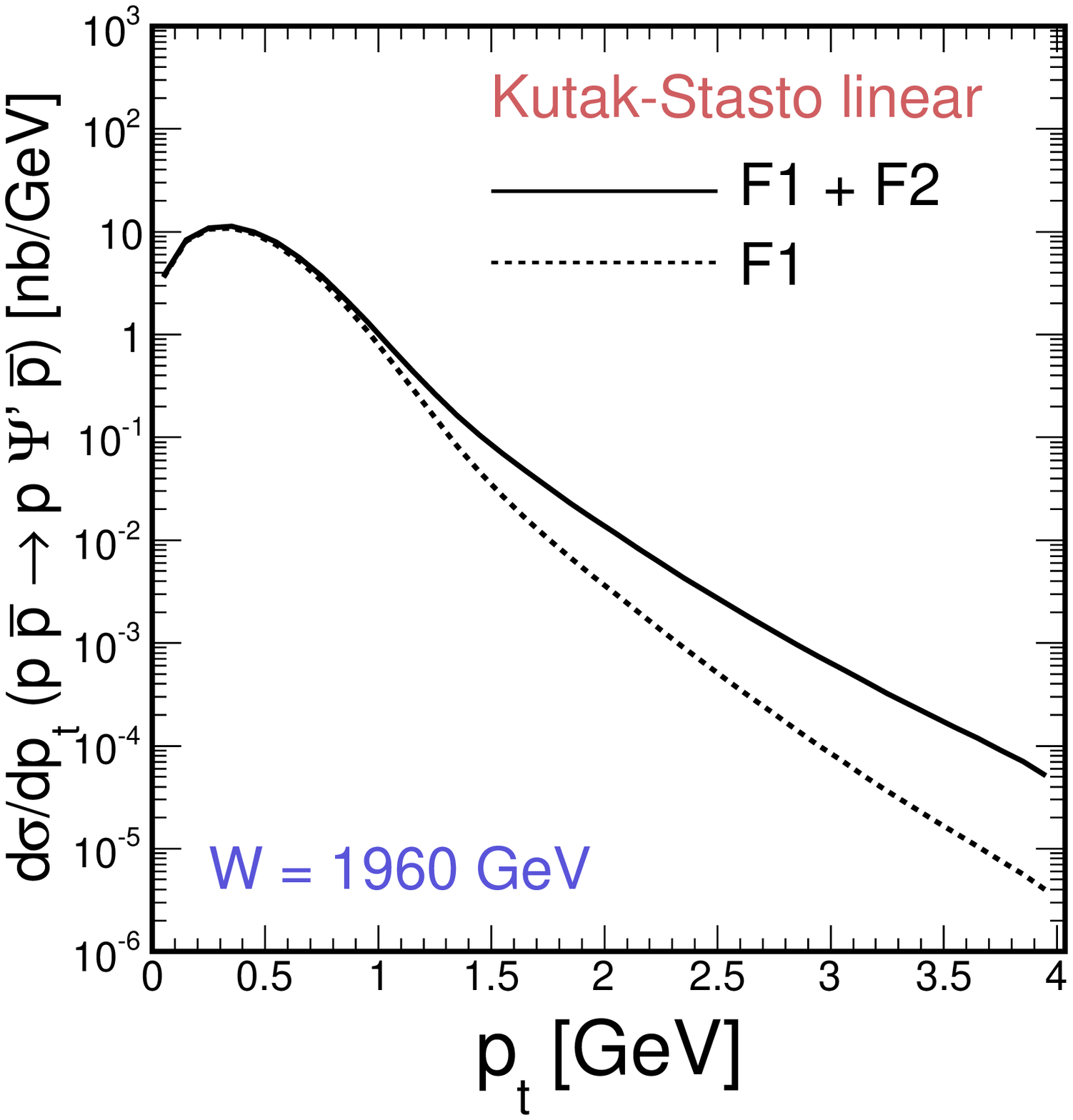}
\includegraphics[width=5.0cm]{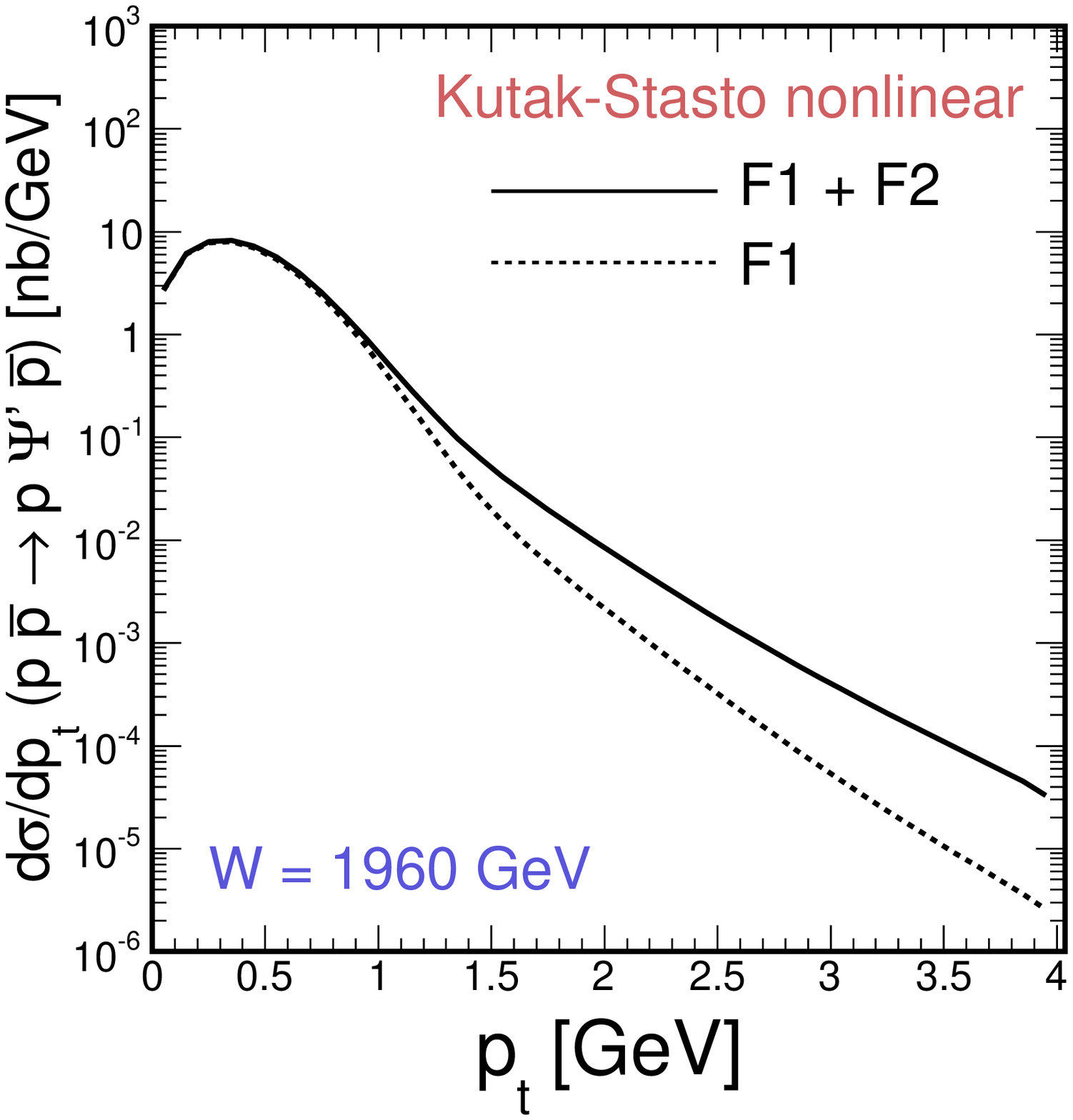}
\caption[*]{ 
$\psi'$ transverse momentum distribution calculated 
in the Born approximation with and without including Pauli
electromagnetic form factor for $\sqrt{s}$ = 1.96 TeV.
\label{fig:dsig_dpt_2S_Born_Tev}
}
\end{center}
\end{figure}

\begin{figure}[!htb] 
\begin{center}
\includegraphics[width=5.0cm]{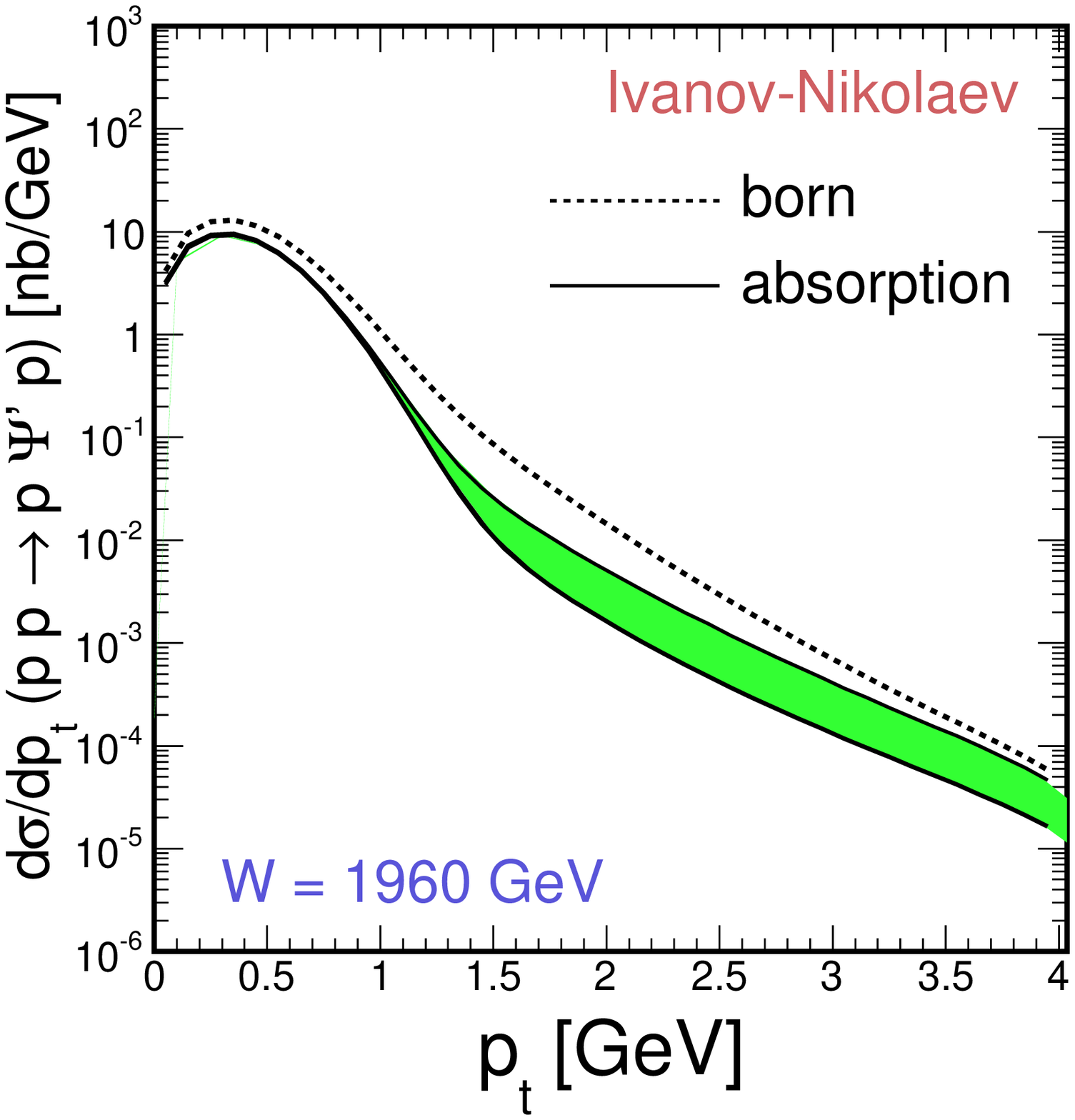}
\includegraphics[width=5.0cm]{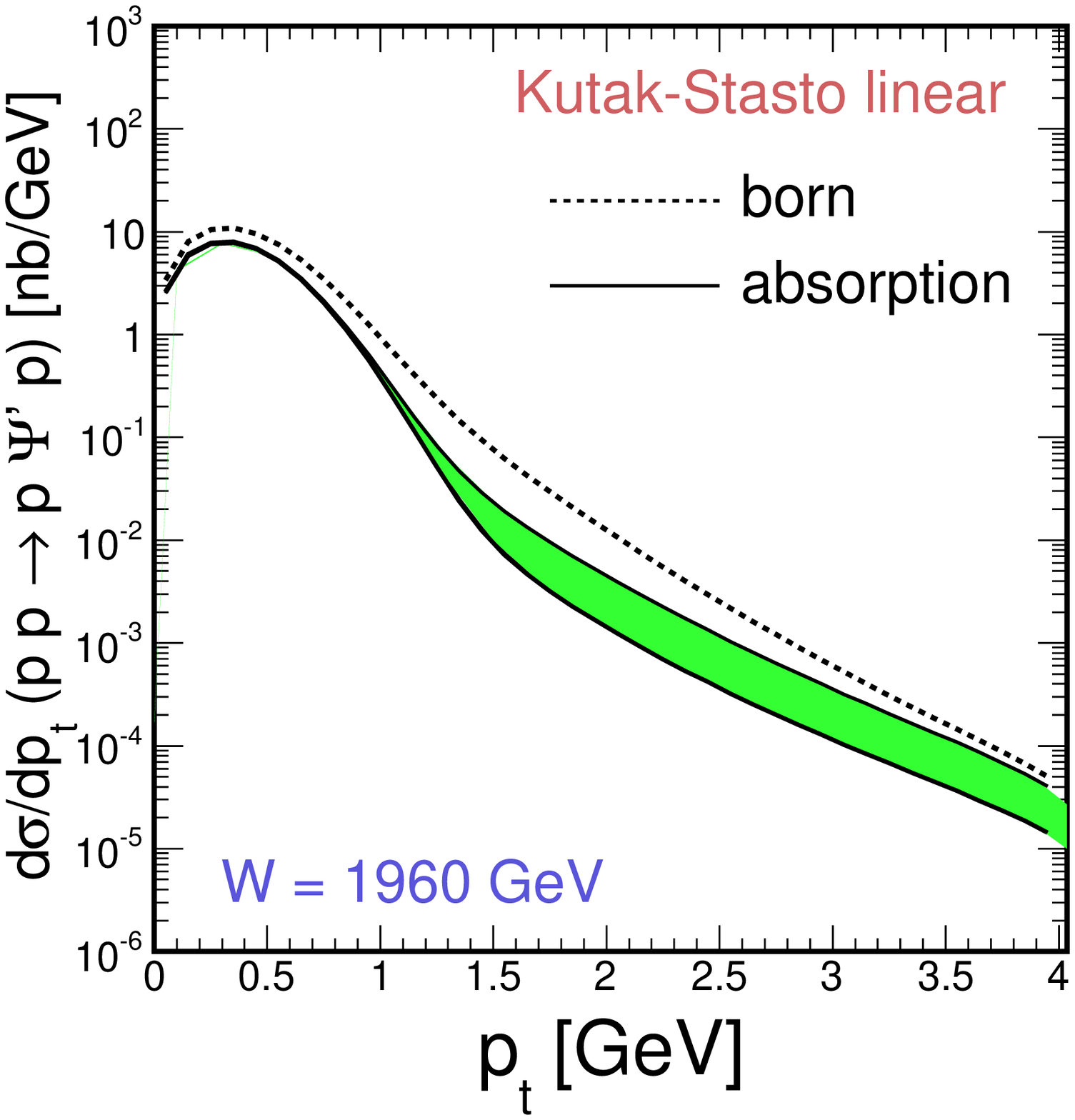}
\includegraphics[width=5.0cm]{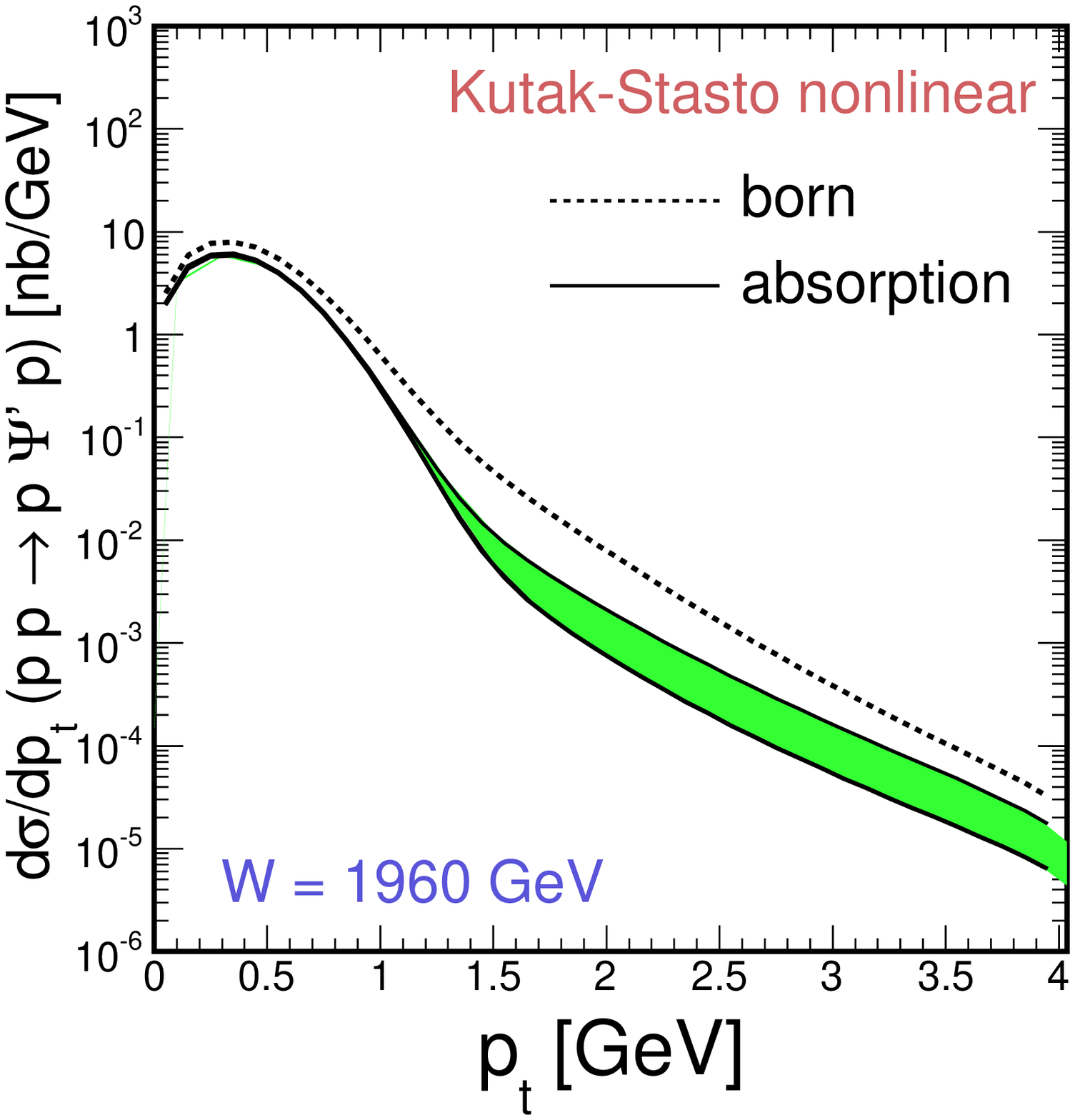}
\caption[*]{ 
$\psi'$ transverse momentum distribution calculated with 
absorption effects (solid line) and in the Born approximation
(dashed line) for $\sqrt{s}$ = 1.96 TeV.
\label{fig:dsig_dpt_2S_absorption_Tev}
}
\end{center}
\end{figure}

The same distributions but for exclusive $\psi'$ production are shown in
Figs.\ref{fig:dsig_dy_2S_absorption_Tev},\ref{fig:dsig_dpt_2S_Born_Tev},
\ref{fig:dsig_dpt_2S_absorption_Tev}.
The shape of the distributions is very similar as for the $J/\psi$ meson.
The corresponding cross section is, however, much smaller.
Again we nicely describe the Tevatron experimental point at
midrapidity.

Summarizing the situation at the Tevatron, we nicely
describe experimental data points at midrapidity both for
exclusive $J/\psi$ and $\psi'$ production which gives further
credibility to our approach.

\section{Conclusions}

In the present paper we have reconsidered exclusive production
of $J/\psi$ meson in the $\gamma p \to J/\psi p$ and
$p p \to p p J/\psi$ reactions within the $k_t$-factorization formalism.
First the total cross section for the $\gamma p \to J/\psi p$ reaction
was calculated as a function of the subsystem energy and confronted with
the HERA data for three different unintegrated distributions from 
the literature.

In comparison to our earlier calculations in the past \cite{SS2007} 
in the present paper we have taken into account both the coupling of photons
via spin-conserving vector coupling with $F_1$ Dirac electromagnetic
form factor as well as spin-flipping tensor coupling with $F_2$
Pauli electromagnetic form factor for the $p p \to p J/\psi p$ and
$p p \to p \psi' p$ reactions.
In distinction to the collinear approach, the theoretical approach used 
here allows to calculate many differential distributions
for the three-body reaction $p p \to p p J/\psi$.
We have calculated not only $J/\psi$ rapidity distribution but also
distributions in $J/\psi$ transverse momentum and distributions
in four-momenta squared $t_1$ or $t_2$. The role of the tensor coupling
with the strength quantified by Pauli electromagnetic form factor
$F_2$. The tensor coupling is important for large $|t_1|$ or $|t_2|$ 
and as a consequence also for large transverse momenta of $J/\psi$.
We have also carefully discussed the role of soft rescatterings
which leads to a shape deformation of all distributions in contrast to
commonly used uniform factor known as gap survival factor.
The uncertainties related to the absorption effects have been discussed.
We have shown that inclusions of inelastic rescatterings leads to further
significant damping of the cross section at the large transverse momenta
of $J/\psi$ and $\psi'$.

Our calculations have been performed for different unintegrated
gluon distributions used previously in the literature. The best agreement 
with the recent LHCb collaboration data has been achieved with UGDF 
which incorporates nonlinear effects in its evolution. This suggests 
an onset of saturation effects, especially for large $J/\psi$
rapidities. Since a simple parametrization of the experimental cross section
for  $\gamma p \to J/\psi$ reaction also leads to a relatively good 
description of the LHCb data no definite conclusion on the onset of saturation
can be drawn.

We have presented our results also for the Tevatron. A good agreement
with the CDF experimental data point at the midrapidity for both 
$J/\psi$ and $\psi'$ has been achieved.

In the future we plan to find a phenomenological UGDF which
simultaneusly describes the $F_2$ deep-inelastic structure function data
and the LHCb collaboration data for semi-exclusive production of
$J/\psi$. Then the photonic-inelastic contributions (the exchanged photon
leaves the remaining system in an excited state) must be taken into
account in the analysis in a similar fashion as recently done for
$\mu^+ \mu^-$ semi-exclusive production \cite{Krakow-Louvain}. 
This clearly goes beyond the scope of the present paper.

\vspace{1.0cm}

{\bf Acknowledgments}

We would like to thank to Ronan McNulty for a discussion of the 
LHCb data.
This work was partially supported by the Polish MNiSW grant 
DEC-2011/01/B/ST2/04535 as well as by
the Centre for Innovation and Transfer of Natural Sciences and 
Engineering Knowledge in Rzesz\'ow.




\begin{thebibliography}{99}

\bibitem{CDF}
  T.~Aaltonen {\it et al.}  [CDF Collaboration],
  Phys.\ Rev.\ Lett.\  {\bf 102}, 242001 (2009).

\bibitem{LHCb_first}
  R.~Aaij {\it et al.}  [LHCb Collaboration],
  J.\ Phys.\ G {\bf 40}, 045001 (2013).


\bibitem{LHCb_second}
  R.~Aaij {\it et al.}  [LHCb Collaboration],
  J.\ Phys.\ G {\bf 41}, 055002 (2014).

\bibitem{Klein:2003vd} 
  S.~R.~Klein and J.~Nystrand,
  Phys.\ Rev.\ Lett.\  {\bf 92}, 142003 (2004). 

\bibitem{SS2007}
W. Sch\"afer and A. Szczurek, Phys. Rev. {\bf D76} (2007) 09014.

\bibitem{MW2008}
L. Motyka and G. Watt, Phys. Rev. {\bf D78} (2008) 014023,
arXiv:0805.2113.

\bibitem{BMSC2007}
A. Bzdak, L. Motyka, L. Szymanowski and J.-R. Cudell,
Phys. Rev. {\bf D75} (2007) 094023.

\bibitem{Goncalves:2011vf} 
  V.~P.~Goncalves and M.~V.~T.~Machado,
  Phys.\ Rev.\ C {\bf 84}, 011902 (2011).

\bibitem{Ducati:2013tva} 
  M.~B.~Gay Ducati, M.~T.~Griep and M.~V.~T.~Machado,
  Phys.\ Rev.\ D {\bf 88}, 017504 (2013).

\bibitem{JMRT2013}
  S.~P.~Jones, A.~D.~Martin, M.~G.~Ryskin and T.~Teubner,
  JHEP {\bf 1311}, 085 (2013).


\bibitem{HERA_new}
  C.~Alexa {\it et al.}  [H1 Collaboration],
  Eur.\ Phys.\ J.\ C {\bf 73} (2013) 2466.

\bibitem{Cisek_phd}
Anna Cisek, PhD Thesis in The Henryk Niewodnicza\'nski Institute of Nuclear
Physics Polish Academy of Sciences, Krak\'ow, Poland, 2012.

\bibitem{INS}
I. P. Ivanov, N. N. Nikolaev and A. A. Savin, Phys. Part. Nucl. {\bf 37}
(2006) 1.

\bibitem{Ivanov:2003iy} 
  I.~P.~Ivanov,
  ``Diffractive production of vector mesons in deep inelastic scattering within k(t) factorization approach,''
  hep-ph/0303053; PhD Thesis, Bonn University.

\bibitem{RSS}
A. Rybarska, W. Sch\"afer and A. Szczurek, Phys. Lett. {\bf B668} (2008)
126.

\bibitem{CSS}
A. Cisek, W. Sch\"afer and A. Szczurek, Phys. Lett. {\bf B690} (2010)
168.


\bibitem{Ryskin}
M.G. Ryskin, Z. Phys. {\bf C57} (1993) 89.


\bibitem{Trawinski:2014msa} 
  A.~P.~Trawi\'nski, S.~D.~G\l azek, S.~J.~Brodsky, G.~F.~de Teramond and H.~G.~Dosch,
  arXiv:1403.5651 [hep-ph].

\bibitem{Nikolaev:1992si} 
  N.~N.~Nikolaev,
  Comments Nucl.\ Part.\ Phys.\  {\bf 21}, 41 (1992).

\bibitem{Kopeliovich:1993gk} 
  B.~Z.~Kopeliovich, J.~Nemchick, N.~N.~Nikolaev and B.~G.~Zakharov,
  Phys.\ Lett.\ B {\bf 309}, 179 (1993).

\bibitem{Nemchik:1996cw} 
  J.~Nemchik, N.~N.~Nikolaev, E.~Predazzi and B.~G.~Zakharov,
  Z.\ Phys.\ C {\bf 75}, 71 (1997).

 \bibitem{CDF94}
CDF collaboration, F. Abe, Phys. Rev. {\bf D50} (1994) 5518.

\bibitem{TOTEM}
TOTEM collaboration, G. Antchev et al., Eur. Phys. Lett. {\bf 101}
(2013) 21002.

\bibitem{H1_a}
H1 collaboration, C. Adloff et al., Phys. Lett. {\bf B541} (2002) 251,
.

\bibitem{H1_b}
H1 collaboration, A. Aktas et al., Eur. Phys. J. {\bf C46} (2006) 585,
.

\bibitem{IN}
I. P. Ivanov and N.N. Nikolaev, Phys. Rev. {\bf D65} (2002) 054004,

\bibitem{KS}
K. Kutak and A. M.  Sta\'sto, Eur. Phys. J. {\bf C49} (2005) 343.

\bibitem{CLSS}
A. Cisek, P. Lebiedowicz, W. Sch\"afer and A. Szczurek, Phys. Rev. {\bf
  D83} (2011) 114004. 

\bibitem{JMRT2013_Psi_2S}
  S.~P.~Jones, A.~D.~Martin, M.~G.~Ryskin and T.~Teubner,
  J.\ Phys.\ G {\bf 41}, 055009 (2014).


\bibitem{Krakow-Louvain}
L. Forthomme, J. Hollar, K. Piotrzkowski, G. de Silveira, 
W. Sch\"afer and A. Szczurek, a paper in preparation.

\end{thebibliography}
\end{document}